%% file: manuscript.tex
\documentclass[%
 reprint,
 amsmath,amssymb,
 aps,
 prl,
 superscriptaddress,
 nofootinbib, floatfix
]{revtex4-2}

\usepackage{caption}

\usepackage{graphicx}
\usepackage{dcolumn}
\usepackage{bm}
\usepackage{xcolor}
\usepackage{tikz}
\usetikzlibrary{quantikz, automata, positioning, arrows}
\usepackage{braket}
\usepackage{placeins}
\usepackage{tcolorbox}
\usepackage[hyperfootnotes=true]{hyperref}
\usepackage[normalem]{ulem}
\usepackage{algorithm}
\usepackage[noend]{algpseudocode}
\usepackage[justification=raggedright,singlelinecheck=false]{caption}
\def\fps@algorithm{tbp}
\makeatother
\def\fps@algorithm{tbp}
\algrenewcommand\algorithmicrequire{\textbf{Input:}}
\algrenewcommand\algorithmicensure{\textbf{Output:}}

\DeclareMathOperator*{\argmax}{arg\,max}
\newcommand{\past}[1]{\overleftarrow{#1}}
\newcommand{\future}[1]{\overrightarrow{#1}}

\definecolor{MYBLUE}{RGB}{0,0,255}
\definecolor{MYRED}{RGB}{255,0,0}

\newcommand{\error}{\epsilon}

\newcommand{\pf}{P_{f|\overleftarrow{x}}}
\newcommand{\rank}{\mathrm{rank}}

\hypersetup{
  colorlinks   = true,    
  urlcolor     = blue,    
  linkcolor    = blue,    
  citecolor    = red      
}

\begin{document}
\preprint{APS/123-QED}
\title{Quantum-Enhanced Inference of Conditional Future Probabilities with Reduced Memory Cost}
\author{Jianjun Chen}
\thanks{Corresponding author}
\email{chen1554@e.ntu.edu.sg}
\affiliation{Nanyang Quantum Hub, School of Physical and Mathematical Sciences,
Nanyang Technological University, Singapore 637371, Singapore}
\affiliation{Centre for Quantum Technologies, Nanyang Technological University, Singapore 639798, Singapore}

\author{Akshat Gupta}
\email{akshat.gupta-4@postgrad.manchester.ac.uk}
\affiliation{Department of Physics \& Astronomy, University of Manchester, Manchester M13 9PL, United Kingdom}
\affiliation{Centre for Quantum Science and Engineering, University of Manchester, Manchester M13 9PL, United Kingdom}

\author{Chengran Yang}
\email{yangchengran92@gmail.com}
\affiliation{Nanyang Quantum Hub, School of Physical and Mathematical Sciences,
Nanyang Technological University, Singapore 637371, Singapore}
\affiliation{Centre for Quantum Technologies, Nanyang Technological University, Singapore 639798, Singapore}

\author{Ariel Neufeld}
\affiliation{Division of Mathematical Sciences, School of Physical and Mathematical Sciences, Nanyang Technological University, Singapore 637371, Singapore}

\author{Jayne Thompson}
\email{jayne.thompson@ntu.edu.sg}
\affiliation{Nanyang Quantum Hub, School of Physical and Mathematical Sciences,
Nanyang Technological University, Singapore 637371, Singapore}
\affiliation{Centre for Quantum Technologies, Nanyang Technological University, Singapore 639798, Singapore}
\affiliation{College of Computing and Data Science, Nanyang Technological University, Singapore 639798, Singapore}
\affiliation{Global Institute of Finance, Technology, and Society (GIFTS), Nanyang Technological University, Singapore 639798, Singapore}

\author{Thomas J.~Elliott}
\affiliation{Department of Physics \& Astronomy, University of Manchester, Manchester M13 9PL, United Kingdom}
\affiliation{Department of Mathematics, University of Manchester, Manchester M13 9PL, United Kingdom}
\affiliation{Centre for Quantum Science and Engineering, University of Manchester, Manchester M13 9PL, United Kingdom}

\author{Mile Gu}
\thanks{Corresponding author}
\email{mgu@quantumcomplexity.org}
\affiliation{Nanyang Quantum Hub, School of Physical and Mathematical Sciences,
Nanyang Technological University, Singapore 637371, Singapore}
\affiliation{Centre for Quantum Technologies, Nanyang Technological University, Singapore 639798, Singapore}

\date{\today}

\begin{abstract}
Many potential future events have devastating consequences, from the onset of earthquakes to financial black swan events. Accurate predictions of their probability of occurrence are vital. Yet as the stochastic systems underlying such events become increasingly complex, these probabilities can depend on ever-growing amounts of past data, which require ever more memory to track. The accuracy of our estimates is ultimately limited by both computational and memory constraints; the former to reduce sampling error, the latter to avoid sampling bias. Traditional quantum algorithms -- such as amplitude estimation -- address only the former, and become untenable especially in near-future quantum devices where quantum memory is a premium. Here, we offer quantum-algorithmic means to reduce systematic bias while maintaining a quadratic speed-up.  Our results thus advance the timeline for the practical deployment of quantum technologies, and illustrate their additional benefits when simulating processes of growing complexity.
\end{abstract}
\maketitle
Given a past sequence of seismic observations, is the probability of a future earthquake 1 or 1.1 in a million? Accurately estimating the conditional probability $p$ of certain future events in stochastic systems can be of vital importance. Conventional means to estimate such $p$ through sampling possible futures are ultimately constrained by the central limit theorem. To produce estimates of $p$ with additive error $\epsilon$, one must draw $\mathcal{O}(1/\epsilon^{2})$ independent samples. Here, quantum amplitude estimation (QAE) promises a quadratic reprieve, reducing that cost to $\mathcal{O}(1/\epsilon)$\,\cite{Brassard2002}. This prospect has fuelled a surge of interest in quantum-enhanced risk analysis and provides a key use-case study often cited in the emerging field of quantum finance~\cite{Woerner2019, Rebentrost2018}.

Yet, realistic deployment of QAE faces a major hurdle: state preparation. QAE methods to estimate such outcome probabilities rely critically on a quantum circuit $\mathcal{A}$ that maps $\ket{0}$ to its quantum sample $\ket{\phi_Z} = \sum_z \sqrt{p_z} \ket{z}$. Consider a process that emits binary symbols \(x_t\) governed by random variable $X_t$ at discrete times. A joint distribution over $L$ time-steps already has \(2^{L}\) possible outcomes, and thus loading their amplitudes by brute force is generally intractable~\cite{sunAsymptoticallyOptimalCircuit2023,zhangQuantumStatePreparation2022}. 

A more feasible approach is to first identify a classical model of the process, then translate this into a quantum circuit gate-by-gate and qubit-by-qubit. However, qubits are much more expensive to control than bits. Meanwhile, classical models of complex processes can be extremely memory intensive -- to the point that they motivated a plethora of classical memory reduction techniques~\cite{chengSurveyModelCompression2020, gholamiSurveyQuantizationMethods2022}. Indeed, there are fundamental theoretical limits on how many bits we must track about a process to faithfully infer its future~\cite{crutchfieldOrderChaos2012, Crutchfield1989}. Drop below this, and prediction bias is unavoidable. With quantum memory being much less readily available, one may anticipate that any reduction in error from QAE will be outweighed by such bias, presenting a major hurdle for practical quantum advantage.

Here, we ask: \emph{Could we simultaneously obtain memory and sampling advantages?} We answer this in the affirmative by leveraging recent advances in quantum-enhanced dimensional reduction~\cite{yang2024dimension}. We illustrate a means for inferring future likelihood that exhibits both advantages: (1) Under memory constraints, they can infer future likelihoods of complex stochastic processes with significantly reduced bias, (2) they inherit the quadratic computational advantage of Quantum-enhanced Monte Carlo methods. Taken together, these advances help accelerate the timeline in which quantum computers can deliver a practical advantage.
\section{\label{sec:pipeline}Framework}

\textbf{Problem Statement}. Consider a dynamical system that emits observable outputs $x_t \in \mathcal{X}$ at each discrete time step $t$. Each $x_t$ is governed by a corresponding random variable $X_t$, such that the system's observable statistics are captured by the bi-infinite sequence $\overleftrightarrow{X}= \dots X_{-2}X_{-1}X_0X_1X_2\dots$. We focus on stationary stochastic processes, in which the joint distribution of any consecutive $L$ outputs $x_{0:L}:= x_0x_1\dots x_{L-1}$ is invariant under time translations. This stationarity allows us to set $t=0$ as the present without loss of generality. Each realisation of the process has a specific past $\past{x}= x_{-\infty:0}$ and a corresponding conditional future $\overrightarrow{x} = x_{0:\infty}$, governed by the distribution $P(\overrightarrow{X}\mid \past{x})$.

Our objective is to estimate the likelihood of a target future event up to some time horizon $L$, formalised as the subset of futures that satisfy some binary property $f$. Mathematically, we define $f$ as a binary function such that $x_{0:L}$ satisfies the property if and only if $f(x_{0:L}) = 1$. Our goal is then to estimate the conditional probability $p$ that a sequence $x_{0:L}$ distributed according to $P(X_{0:L}\mid \past{x})$ satisfies $f$, that is, to compute $p := P[f(x_{0:L}) = 1 \mid \past{x}] := P_{f\mid \past{x}}$.

\textbf{Classical Baseline}. In classical settings, a standard approach to estimating the probability of a target event is statistical sampling using a predictive model $\mathcal{M}$. Formally, such a predictive model operates by encoding each past $\past{x}$ into a memory state $s_i \in \mathcal{S}$. At each time step, the dynamics of the model are then specified by transition probabilities $T^x_{ij} = P(s_j, x \mid s_i)$, describing the probability that it emits $x$ and transitions to state $s_j$ when in state $s_i$. A model $\mathcal{M}$ is then said to faithfully represent a particular process $P$ if iterating this procedure for $L$ time steps results in a sequence drawn from the conditional distribution $P(X_{0:L} \mid \past{x})$. Such models can be distilled from recurrent neural networks and transformer models, or alternatively constructed analytically based on an underlying understanding of the process we wish to sample.

Here, we refer to the process of generating one sample from $P(X_{0:L} \mid \past{x})$ as one simulation. By running this simulation $N$ times, the proportion of samples that satisfies $f$ produces an estimate $\widehat{p}$ of $p$. The performance of any such estimation algorithm can then be benchmarked by the root mean squared error (RMSE):
\begin{align}
    \epsilon = \sqrt{\mathbb{E}\!\left[(p-\widehat{p})^2\right]}, \label{eqn:standard_error}
\end{align}
where the expectation is taken over the distribution of $\widehat{p}$.

There are then two main resource concerns. The first is the number of times, $N$, that we need to simulate the process. This scales as $1/\epsilon^2$ to reach a designated error threshold $\epsilon$~\cite{cerouAdaptiveMultilevelSplitting2019, bucklewIntroductionRareEvent2004, glassermanMonteCarloMethods2010}. The second is the number of distinct memory states $D = |\mathcal{S}|$ tracked by the model, which can grow rapidly in modern sequence-modelling settings.

\begin{figure*}[ht]
    \centering
    \includegraphics[width=0.85\linewidth]{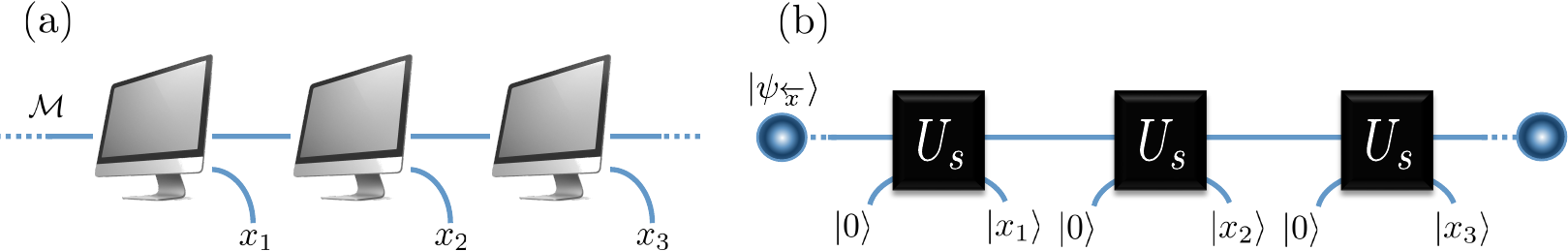}
    \caption{\textbf{Classical vs. Quantum Models}. (a) A classical model of a stochastic process operates by tracking relevant past information in its memory states $\{s_j\}$, encoded within the memory system $M$. At each time step, a classical model in state $s_j$ emits a symbol $x$ and transitions to some state $s_k$ with probability $T^{x}_{jk}$. (b) A quantum model consists of a quantum memory initialised in a state $\ket{s_j}$, based on the past observations $\protect\past{x}$. At each time step $t$, the unitary $U_s$ entangles the memory register with a fresh ancillary system, imprinting the output $\ket{x_t}$ onto the ancilla and simultaneously updating the memory. The output $\ket{x_t}$ can be measured immediately to generate a sequence for the stochastic process defined by the unitary $U_s$ or left coherent in the quantum circuit for further processing.\label{fig:unitary_simulator}}
\end{figure*}

\textbf{Quantum Amplitude Estimation (QAE)} offers a quadratic reduction in the number of simulations needed. It assumes that we begin with $\mathcal{A}$, a state-preparation circuit that generates a quantum sample of $X_{0:L}$, i.e., $\sum_{x_{0:L}} \sqrt{P(x_{0:L}|\past{x})}\ket{x_{0:L}}$. We can then define the Grover operator 
\begin{align}
    \mathbf{Q} = -\mathcal{A}S_0\mathcal{A}^\dagger S_\chi,
\end{align}
where, $S_{0} = \mathbb{I} - 2\ket{0}\bra{0}$ is a reflection operator about the all-zero state, and $S_{\chi}$ is the oracle that marks the target subspace, i.e., $S_\chi = \mathbb{I} -2\sum_{x_{0:L}:f(x_{0:L})=1}\ket{x_{0:L}}\bra{x_{0:L}}$, where $f(x_{0:L})=1$ if $x_{0:L}$ satisfies the property $f$ and $0$ otherwise~\cite{grover1996fast}. Standard QAE uses the Grover operator $\mathbf{Q}$ within a phase estimation subroutine, such that the number of times we need to execute $\mathcal{A}$ scales only as $1/\error$~\cite{Brassard2002}. Subsequently, the limitations of present-day quantum technology motivated shallow circuit variants that avoid phase estimation via classical post-processing~\cite{Suzuki2020, Grinko2021, Giurgica-Tiron2020, Aaronson2020}.

While QAE provides a quadratic reduction in sample complexity, it offers little relief regarding the cost of generating each sample (i.e., executing $\mathcal{A}$). Given that the number of outcomes of $X_{0:L}$ scales exponentially with $L$, direct synthesis of $\mathcal{A}$ generally requires an exponential number of quantum gates or ancilla qubits~\cite{sunAsymptoticallyOptimalCircuit2023, zhangQuantumStatePreparation2022}. A standard workaround is to leverage an existing classical model $\mathcal{M}$. In principle, if $\mathcal{M}$ can generate classical samples efficiently, then we can use it to construct an efficient $\mathcal{A}$ with similar gate complexity, provided we are willing to commit at least one qubit for each bit. In practice, this often remains impractical. Classical bits are cheap, but qubits are not. Tracking a dozen floating-point numbers to 32-bit precision is trivial on today’s computers, yet would quickly saturate the capabilities of a near-future quantum device.

\textbf{Model Reduction and Distortion}. Given the above constraints, model reduction is paramount. The simpler our model $\mathcal{M}$, the easier it will be to implement $\mathcal{A}$. This motivates the question: what is the minimal memory -- in terms of memory dimension $D$ (that is, how many distinct configurations $\mathcal{M}$ must hold) -- required to sample $P(x_{0:L}\mid \past{x})$ faithfully? By identifying such a representation, we can encode the past into $D$ distinct states, $\mathcal{S} = \{s_k\}_{k=1}^{D}$, using only $O(\log_2 D)$ bits of memory.

This question is extensively studied in complexity science. Every stochastic process $\mathcal{P}$ possesses an intrinsic measure of complexity given by $D_c$, which lower-bounds the memory dimension of $\mathcal{M}$. Its logarithm, $d_c = \log_2(D_c)$, is known as the \emph{topological state complexity} of $\mathcal{P}$~\cite{Crutchfield1989} (see Appendices \ref{app:e-machine} and \ref{app:complexity-measures} for details). The corresponding memory states $\{s_k\}$ are known as \emph{causal states}, capturing the minimal belief states an agent must retain in memory about the past for optimal prediction of the future -- and arising naturally within contemporary transformer and recurrent neural networks~\cite{shaiTransformersRepresentBelief}.

Yet, even when we can build such minimal models, their memory requirements may be immense or unbounded -- especially when future behaviour is sensitive to continuous parameters or involves long-range temporal correlations~\cite{Elliott2020, marzenInformationalCausalArchitecture2015}. This necessitates approximate models, such as clustering and finite-precision encodings~\cite{chengSurveyModelCompression2020, choiLimitNetworkQuantization2017, gongCompressingDeepConvolutional2014, marzenEvolutionLossyCompression2017}, in which the memory dimension is constrained to some desired limit $\tilde{D} < D_c$. This memory reduction results in distortion, such that our approximate model estimates $\hat{p}$ by sampling from some induced distribution $\tilde{P}(X_{0:L}\mid \past{x})$ that approximates $\mathcal{P}$. The resulting errors in estimating $p$ then comprise (1) random contributions $\epsilon_{\mathrm{rand}}$, which reflect errors due to a finite number of samples, and (2) systematic bias $\epsilon_b$, arising from inherent distortions in the model. Crucially, the latter does not diminish with $N$, and as $N \to \infty$, our total error satisfies $\epsilon \to |\tilde{P}_{f\mid \past{x}} - P_{f\mid \past{x}}|$, which does not approach $0$. We therefore have a fundamental trade-off between reducing memory cost and reducing predictive error.

\section{Approach}

\textbf{Algorithm Outline}. Our aim is to further reduce the resources required to synthesise a suitable state-preparation operator $\mathcal{A}$ below what is possible through conventional bit-to-qubit translation. Our approach begins with a target process $\mathcal{P}$ that has topological state complexity $\log_2 D_c$. A memory-minimal classical model $\mathcal{M}_c$ that generates predictions with zero bias then has $D_c$ memory states. Meanwhile, enforcing a target memory constraint $\tilde{D} < D_c$ would naturally force the use of a biased model, and any $\mathcal{A}$ obtained by translating such a model would inherit similar bias. Our approach instead exploits the fact that quantum models with a fixed memory dimension $\tilde{D}$ are more expressive~\cite{Elliott2020, yang2024dimension}, allowing us to design an efficient implementation of $\mathcal{A}$ whose bias is significantly lower than that of the baseline approach.

\begin{figure*}[ht]
    \centering
    \includegraphics[width=0.9\linewidth]{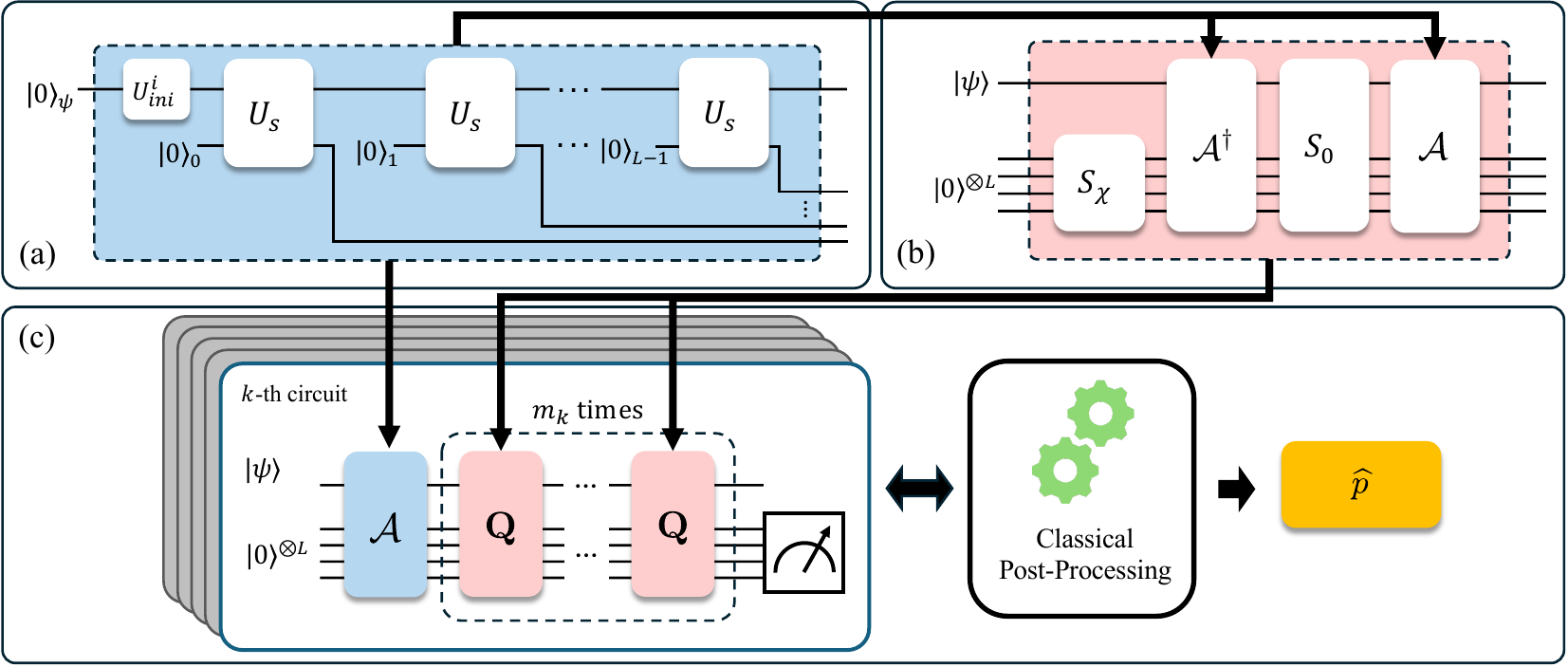}
    \caption{\textbf{Algorithm Schematic}. Our method for memory-reduced conditional probability estimation tackles the data-loading problem through the use of a recurrent quantum circuit $\mathcal{A}$ (a). Through $L$ repeated applications of $U_s$ on a suitably initialised memory system and $L$ ancilla qubits, the circuit is able to generate a quantum superposition of all possible futures up to a time horizon of $L$. Provided the memory system (top line) is properly accounted for, we can build a suitable Grover iterator $\mathbf{Q}$ (b). We can then construct a family of quantum circuits $N_k$, each involving $m_k$ repeated applications of $\mathbf{Q}$. The resulting measurement statistics from these circuits then allow us to estimate $\hat{p}$ (c). For cases where $f$ is an indicator function that flags target strings, $S_\chi$ can be constructed from a multi-controlled single-qubit gate. The reflection operator $S_0$ can also be constructed using a multi-controlled single-qubit gate. 
    }
    \label{fig:comprehensive_algorithm}
\end{figure*}

Let $\mathcal{M}_c$ have memory states $\{s_j\}$ and transition elements $T^{x}_{jk}$. We can then define an associated quantum model $\mathcal{M}_q$ that associates each $s_j$ with a corresponding quantum memory state $\ket{s_j}$~\cite{Binder2018}. Thus, if $\mathcal{M}_c$ encodes $\past{x}$ in state $s_j$ within memory $M$, its associated quantum model encodes $\past{x}$ in $\ket{s_j}$ within a quantum memory $M_q$. At each time step $t$, the quantum model operates by introducing a fresh ancilla initialised in $\ket{0}$ and enacting a unitary $U_s$ satisfying
\begin{align}
    U_s \ket{s_i}\ket{0} = \sum_{j=0}^{|\mathcal{S}|}\sum_{x\in \mathcal{X}} \sqrt{T^x_{i,j}} \ket{s_j}\ket{x}, \label{eqn:unitary_operation}
\end{align}
which couples its memory to the ancilla (see Figure~\ref{fig:unitary_simulator}). Here, there is no requirement that the set $\{\ket{s_j}\}$ be linearly independent~\cite{Binder2018}. As such, the associated memory dimension of $\mathcal{M}_q$, denoted $D_q$, can be strictly less than $D_c$ (See Appendix \ref{app:physical-mechanism}). When this occurs, it immediately enables a memory advantage in the zero-bias regime. More generally, when given a target memory $\tilde{D} < D_c$, we have a means to design quantum models whose conditional future statistics more closely approximate $\mathcal{P}$ (see Algorithm 4 in Appendix \ref{app:quantum-dimensional-reduction}).

We then leverage the resulting quantum model to construct a Grover operator. First, we need to define its components: the state-preparation operator $\mathcal{A}$ and the reflection operators $S_\chi$ and $S_0$. For a future sequence of length $L$, i.e. $x_{0:L}=x_0x_1\dots x_{L-1}$, the quantum circuit will consist of a memory register of size $n = \lceil \log_2 |\tilde{D}|\rceil$ qubits and $L$ output registers, each of size $\lceil\log_2|\mathcal{X}|\rceil$ qubits. This initial circuit can be represented as $\ket{0}_\psi\ket{0}_0\ket{0}_1\dots \ket{0}_{L-1}$, where $\ket{0}_\psi$ is the memory register and $\ket{0}_i$ is the output register to store the $i$-th index of the sequence $x_{0:L}$, each initialized in the computational zero state. 

To construct the state-preparation operator, we first initialise the memory register in the appropriate $\ket{s_j}$ consistent with the observed $\past{x}$. This can always be done via some unitary $U_{\mathrm{ini}}$ (see Appendix \ref{appendix:creation_ini}). Subsequently, for each symbol of the sequence, the unitary $U_{s}$ is applied to the memory register and a new output register initially in the zero state. This sequence of operations defines the state-preparation operator $\mathcal{A}$ as
\begin{equation}
    \mathcal{A} = U_{s}^{(\psi,L-1)}\dots U_{s}^{(\psi,1)}U_{s}^{(\psi,0)}U_{\mathrm{ini}},
    \label{eqn:A_equation}
\end{equation}

where $U_{s}^{(\psi, i)}$ denotes the application of $U_{s}$ on the memory register and the $i$-th output register. Application of $\mathcal{A}$ results in $\sum_{x_{0:L}} \sqrt{P(x_{0:L}|\past{x})}\ket{x_{0:L}}\ket{s_{\past{x}'}}$ where $\past{x}' = \past{x}x_{0:L}$. It thus does not strictly prepare the q-sample state, but rather the q-sample entangled with a memory state for generating samples beyond time $t = L$. 

In Figure~\ref{fig:comprehensive_algorithm}, we illustrate that provided this memory is properly accounted for, the principles of QAE still apply. Specifically, in constructing the Grover operator, the oracle gate $S_\chi$ is applied only to the output registers as per standard QAE. However, the rest of the components, i.e., the operator $\mathcal{A}$, its reversal $\mathcal{A}^\dag$ and the reflection operator $S_0$ are applied to both memory and output registers.

An outline of the resulting end-to-end algorithm is documented in Algorithm~\ref{alg:algorithm}. This algorithm takes as input (i) a classical model $\mathcal{M}_c$ of the process $P$, (ii) the binary indicator function $f$, (iii) the number of future time-steps we wish to sample $L$, (iv) the past $r$ outputs observed, and (v) the available memory dimension $\tilde{D}$. The algorithm then outputs an estimate of $\pf$ with both reduced variance and bias. Here, the subroutine \texttt{QuantumModel} begins by transforming a supplied $\mathcal{M}_c$ to a quantum model that replicates its statistical behaviour exactly. The \texttt{Truncate} subroutine then provides a means to reduce the memory of this model to dimension $\tilde{D}$. We can then connect these models with a variant of quantum amplitude estimation, denoted by \texttt{QAE} ~\cite{Brassard2002, Grinko2021, Giurgica-Tiron2020, Aaronson2020}, that can leverage such quantum models to estimate probabilities of future events up to some future time-horizon $L$ (See Figure~\ref{fig:comprehensive_algorithm}).

\begin{algorithm}[H]
\caption{(MRQPE)\\Memory-reduced Quantum Probability Estimation\label{alg:algorithm}}
\begin{algorithmic}[1]
\Require 
\Statex $\mathcal{M}_c = (\mathcal{S},T^x_{ij})$: Classical model of a target process $\mathcal{P}$.
\Statex $f$: binary function that marks the target subspace.
\Statex $L$: length of future sequence.
\Statex $x_{-r:0}$: past sequence of length $r$.
\Statex $\tilde{D}$: desired number of states.
\Ensure Probability estimate $\widehat{p}$
\State $\{K^x\}, U_{s} $ $\gets $ \Call{QuantumModel}{$\{T^x\}$} (See Appendix \ref{appendix:Ustep_operator} Algorithm 2)
\If{$\tilde{D}< |\mathcal{S}|$}
    \State $\{K^x\}, U_{s} \gets$ \Call{Truncate}{$\{K^x\}, \tilde{D}$} (See Appendix \ref{app:quantum-dimensional-reduction} Algorithm 4)
\EndIf 
\State $U_{\mathrm{ini}} \gets$ \Call{EncodePast}{$x_{-r:0}$, $\{K^x\}$} (See Appendix \ref{appendix:creation_ini} Algorithm 3)
\State Construct operator $\mathcal{A}$ from $U_{\text{ini}}$ and $U_{s}$ (Eq.~\eqref{eqn:A_equation})
\State Construct oracle $S_\chi$ to implement binary function $f$: $f(\ket{x_0x_1\dots x_{L-1}})=\begin{cases}-\ket{x_{0:L}} &\text{ if } f(x_{0:L}) = 1 \\ \ket{x_{0:L}} &\text{otherwise} \end{cases}$
\State Construct Grover operator $\mathbf{Q} = -\mathcal{A} S_0 \mathcal{A}^\dagger S_\chi$
\State Estimate $\widehat{p} \gets$ QAE$(\mathbf{Q}, \mathcal{A})$
\State \Return $\widehat{p}$
\end{algorithmic}
\end{algorithm}

\begin{figure}[tbh]
    \centering
    \includegraphics[width=0.6\linewidth]{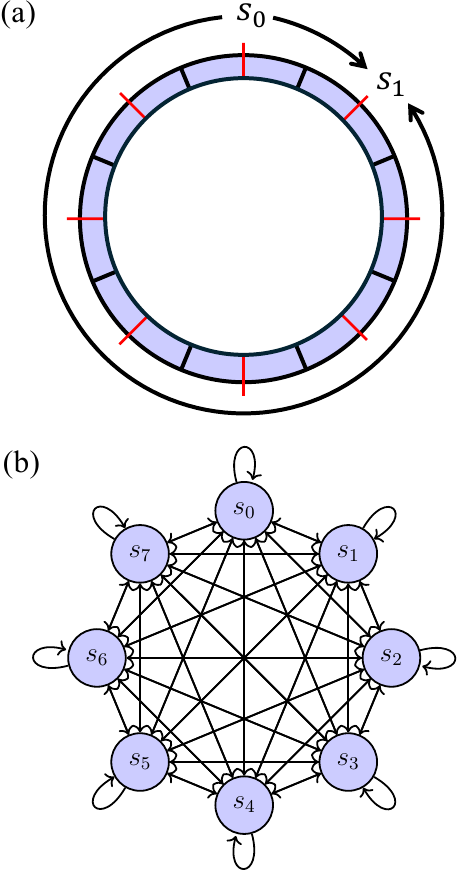}
    \caption{\textbf{Cyclic random walk and its discretisation}. (a) We can discretise a cyclic random walk by dividing the unit interval into equal segments, each represented by its midpoint $s_j = j/N, \ j=0,1,\dots, N-1$ (red markers). Between any two states, there exist two possible transition pathways — one clockwise and one anticlockwise. (b) Given a continuous position $X_t$, the discrete state of the system at time $t$ is defined as the closest segment midpoint $s_j$, denoted $S_t=s(X_t)$. The transition probability from $s_k$ to $s_j$ is then $P(s_k|s_j) = P(X^{t+1} \in \mathcal{I}_k|X^t=x_j)$, where $\mathcal{I}_k = \left\{x:\left|x-\frac{k}{N}\right|<\frac{1}{2N}\right\}$ denotes the interval associated with the segment $s_k$. Here, we consider the $8$-state discretisation as the target process (see Appendix~\ref{app:cyclic-random-walk} for more details).}
    \label{fig:cyclic-random-walk}
\end{figure}

\textbf{Complexity.} As two-qubit operations are generally costly to implement, we focus on the number of CNOT gates required to realise the Grover operator. For a model of memory dimension $D$, each $U_{s}$ acts on $\lceil \log_2 |D|\rceil$ memory qubits and $\lceil \log_2 |\mathcal{X}|\rceil$ output qubits. Thus, simulating $L$ time steps yields a total CNOT gate complexity for implementing $\mathcal{A}$ of $\mathcal{O}(L\cdot D^2\cdot |\mathcal{X}|^2)$. In the scenario of zero bias, a quantum model would have CNOT gate complexity of $\mathcal{O}(L\cdot D_q^2\cdot |\mathcal{X}|^2)$. In contrast, if we were to directly translate the classical model of memory dimension $D_c$ into a quantum circuit, then the memory would be of dimension $D_c$, resulting in a CNOT gate complexity of $\mathcal{O}(L\cdot D_c^2\cdot |\mathcal{X}|^2)$ (see Appendix~\ref{append:gate_complexity} for more details). 

This distinction can be very significant. Even when starting with a classical model that achieves a theoretical minimal memory dimension $D_c$, $D_q$ can be significantly lower. There exists families of $\mathcal{P}$ where $D_q$ remains bounded while $D_c$ scales without bound~\cite{thompsonCausalAsymmetryQuantum2018a, Elliott2020, wuImplementingQuantumDimensionality2023}. In the illustrative example below, we see that setting $\tilde{D} < D_q$ can result in quantum models whose predictions are significantly more accurate than classical counterparts.

\begin{figure*}[t]
    \centering
    \includegraphics[width=0.9\linewidth]{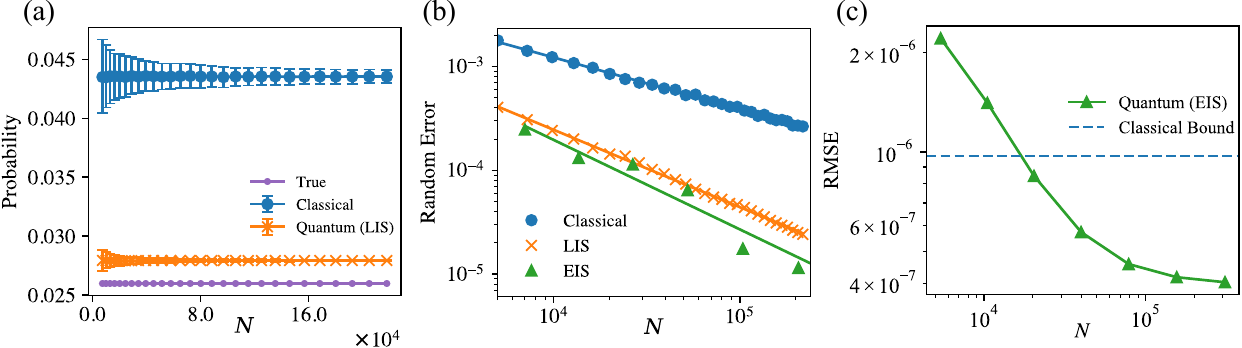}
    \caption{\textbf{Illustration of quantum advantage on a cyclic random walk}. Here we compare the performance of classical Monte Carlo methods to our memory-reduced quantum probability estimator (MRQPE) when simulating a $8$-segment cyclic random walk when memory dimension is constrained to $4$. (a) illustrates the estimates of the probability $p = P(x_{0:3} = 100|x_{-3:0}=000)$ using classical Monte Carlo (blue), MRQPE with linear (LIS) scheduling (see Appendix~\ref{app:MLAE} for details on LIS and EIS scheduling) as a function of the number of simulations $N$, with the true value of $p$ plotted in comparison (purple). The error bars represent the error bars corresponding to the 90th and 10th percentiles over 1000 independent trials. (b) illustrates the advantageous scaling in random error for MRQPE using LIS (orange) and EIS schedules (green), when compared to classical baseline Monte Carlo methods (blue). The numerical values of the fitting parameters and the corresponding correlation coefficient $r^2$ can be found in table~\ref{tab:error_analysis} in Appendix \ref{appendix:convergence_results}. (c) illustrate the average RMSE of our MRQPE method (with memory dimension $\tilde{D} = 4$) when estimating events with probabilities from $10^{-9}$ to $10^{-8}$ over 100 independent trials (green) as a function of $N$. This is compared against the classical bound (blue), representing the RMSE of classical Monte-Carlo methods with identical memory constraints, in the limit of infinite samples (i.e., $N \rightarrow \infty$). More details on the simulation hyperparameters are given in Appendix~\ref{appendix:hyperparameters}.}
    \label{fig:cyclic_random_walk_results}
\end{figure*}

\section{Illustrative Example}

We illustrate the efficacy of our methods in a number of settings. Here, we outline the case of a particle undergoing a stochastic random walk on a ring. In the Appendix~\ref{app:examples}, we include several other examples exhibiting similar advantage, including non-Markovian processes such as the discretised autoregressive models (Appendix~\ref{app:ar2-model}). 

Consider now a particle on a ring~\cite{garnerUnboundedMemoryAdvantage2017}. Let $x_{t} \in [0,1)$ denote the position of the particle at time step $t$ on a ring. At each time step,
\begin{align}
    x_{t+1}= \mathrm{Frac}(x_{t} + y),
\end{align}
where the random variable $y\sim P(Y)$, describes the particle's stochastic movement in space, and $\mathrm{Frac}(z)=z-\lfloor z \rfloor$ maps this to the topology of the ring. Here, we take $P(Y)$ to be a zero-mean Gaussian with variance $\sigma^2 = 0.01$. The resulting sequence $\overleftrightarrow{X}=\dots X_{-2}X_{-1}X_0X_1X_2\dots$ forms a continuous-valued discrete-time stochastic process, which models ring-based diffusion dynamics that can appear in diverse contexts such as ring networks and phase noise~\cite{zhong2005non, doi:10.1073/pnas.2210622120,piemonteseDiscreteTimeModelsPerformance2024}.

Exact simulation of this process would require tracking $[0,1)$ to perfect precision. Thus, in digital simulation, the standard practice is to consider it satisfactory to track $x_t$ some to $k$ bits of precision~\cite{gray1998}. We take this as our target process. For proof of concept, we first take $k = 3$, and will late return to discuss how our results generalise for higher values of $K$. The target process then involves dividing the ring into $N=2^k=8$ equal segments (see Figure \ref{fig:cyclic-random-walk}), and thus represents a $8$-state process.

We ask: if memory is constrained to four-state models ($\tilde{D} = 4$); how well could we infer conditional futures on the target $8$-state process? We compare the performance between (a) baseline classical Monte-Carlo methods using the optimal classical model of dimension $\tilde{D}$ (as measured by KL divergence rate~\cite{yangMeasuresDistinguishabilityStochastic2020}) and (b) our dimension-reduced quantum probability estimator (Algorithm 1).

\begin{figure}[tbh]
    \centering
    \includegraphics[width=0.7\linewidth]{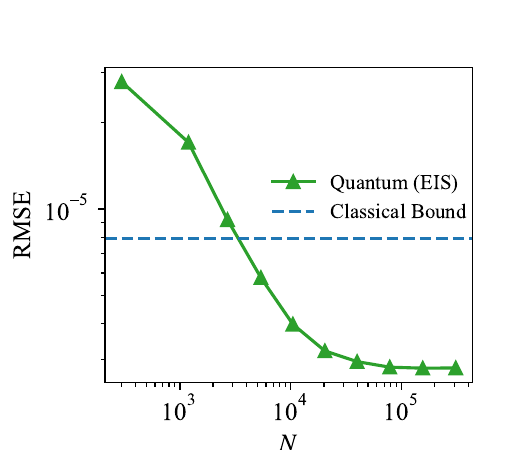}
    \caption{\textbf{Quantum advantage when simulating a 16-state cyclic random walk.} Here, we consider the average RMSE error when estimating all events with probabilities from $10^{-9}$ to $10^{-8}$ for the 16-state cyclic random walk (see Appendix~\ref{app:list-of-futures} for details). As in the the 8-state case, both classical baseline and quantum models are set to have memory dimension of $4$. As in Figure~\ref{fig:cyclic_random_walk_results}c, the blue line then denotes the the lower bound on RMSE achieved using classical models --- even if we allowed unlimited samples. Meanwhile the green line plots the RMSE error of our quantum models as function of number of simulations. We see that quantum models can achieve RMSE below classical limits using only around 3000 samples.}
    \label{fig:16state}
\end{figure}

Figure~\ref{fig:cyclic_random_walk_results}a illustrates the dual advantage of our quantum methods in mitigating sampling error and bias in the context of estimating the probability of observing the output sequence $x_{0:3}=100$ conditioned on a particular past (Here $x_{-3:0}=000$). In addition to the standard quadratic-speed-up due to reduced sampling complexity (Figure~\ref{fig:cyclic_random_walk_results}b), the quantum model exhibits significantly less bias -- converging to values significantly closer to the true probability. To demonstrate the typicality of the bias advantage, we then consider the set of all 36 events whose actual probability of occurrence conditioned on the same past lies between $10^{-9}$ and $10^{-8}$ (see Appendix~\ref{app:list-of-futures} for details). The resulting bias in our classical baseline $4$-state model results in an RMSE of $\sim 10^{-6}$ when averaged across all 36 events -- even in the limit of infinite samples (Figure~\ref{fig:cyclic_random_walk_results}c). In contrast, the average RMS error of our quantum models dips below this bound at around $N\sim 2\times10^4$ samples. Such an advantage would be immediately beneficial for rare-event classification, where we wished to determine that the probability of a target adverse outcome is less than some $\epsilon \ll 1$.

This quantum advantage is not limited to a particular choice of precision $k$ for the target process. Figure~\ref{fig:16state} documents the analogue of Figure~\ref{fig:cyclic_random_walk_results}c for the case where the target process aims to simulate the dynamics of $x_t$ to $k=4$ bits of precision, and thus requires a memory dimension of $N = 16$ states to model exactly. We then continue to constrain both the classical baseline and quantum models to a memory dimension of $4$. We see that the average RMSE of the classical models on the rare events (with probabilities between $10^{-9}$ and $10^{-8}$) remains bounded below by $8\times10^{-4}$ in the limit of infinite samples. Meanwhile, a $4$-level quantum model can undercut this classical bound with approximately $3000$ samples --- more than a factor of $6$ fewer than required for the $k=3$ baseline. Indeed, in Appendix~\ref{app:finer-discrete}, we outline why we expect this quantum advantage to persist for increasing values of $k$ as we approach the continuum limit.

\section{Discussion}
Quantum amplitude estimation promises quadratic speed-ups in estimating the likelihood of future events, but its practical application faces a significant hurdle: an efficient \emph{quantum sampler} to generate such future distributions in quantum superposition. Here, we provide a means to mitigate this issue when inferring future probabilities of general stochastic processes. Our resulting quantum-enhanced future inference algorithm exhibits (i) the typical speed-up of quantum amplitude estimation from a quadratic improvement in how many times we need to simulate the process and (ii) further improved memory-distortion trade-offs that enable reduced errors from finite-memory constraints. The two advantages synergize - providing potential quantum advantage beyond quadratic speed-up. Our procedure is broadly applicable. It requires as input only a classical predictive model of a process. Such a model could be obtained in a number of established methods, from base analytical models (e.g. Markov machines, random walks) to machine-learned models (e.g. RNNs, Transformers, together with classical model reduction techniques). 

Here, we have primarily focused on memory as a key limiting factor for the first generation of universal quantum computing devices, as we anticipate that such devices will be the first to support practical-scale quantum Monte Carlo methods. However, proof-of-principle experimental demonstrations may be possible on nearer-term noisy quantum devices. Indeed, dimensionally reduced single-qubit quantum models have already been experimentally realised~\cite{ghafari2019dimensional}, including those exhibiting provable accuracy advantage~\cite{wuImplementingQuantumDimensionality2023}. Meanwhile, our preliminary analysis indicates that dimensional reduction could provide further benefits in such noisy settings. In particular, the size of our unitary $U_s$ depends on the available quantum memory: the larger $U_s$ is, the more elementary gates are required for its implementation, with each gate introducing additional noise. Thus, the additional bias introduced by dimensionally reducing the memory may be more than offset by the benefits of requiring fewer gates. In Appendix~\ref{app:noisy_proc}, we perform an analysis assuming noise levels comparable to those of current Quantinuum processors when sampling from the target $8$-state cyclic random walk considered above. Our results show that the overall bias in estimating conditional future probabilities generally \emph{decreases} when we use a $2$-qubit dimensionally reduced quantum model to simulate the target process, compared with the exact $3$-qubit model.

There are also several promising future directions. One is to incorporate classical importance sampling techniques. There, sophisticated methods developed allow the potential to pre-process a classical model to produce a variant that samples more often from atypical sequences -- and thus accelerate estimating their probability of occurrences~\cite{bouchetRareEventSampling2019, tokdar2010importance}. Quantum processing turns out particularly effective in reducing the complexity of such variants, where resulting models can have unbounded memory advantage~\cite{aghamohammadi2018extreme}. Incorporating such pre-processing techniques could further enhance our quantum algorithm's capability of estimating the likelihoods of rare events. A second is further circuit optimisation. We note that the gate complexities quoted for our quantum models are presently worst-case scenarios. In practice, there is significant freedom in choosing the interaction unitary: as the stochastic model constrains only its action when the output register is initialised in $\ket{0}$~\cite{Binder2018}; a freedom that may enable further reduction in circuit complexity~\cite{iten16}. Meanwhile, integration of our methodologies with existing transformer and RNN models has exciting prospects. Here, the exponential growth in memory costs of present AI models has motivated model quantization -- a classical technique where continuous properties tracked by the AI model are represented by low-bit approximations~\cite{alom2018effective}. Quantum models have shown provably unbounded memory advantage in such settings~\cite{PhysRevX.12.011007}, presenting an ideal case study for replacing classically `quantized' models with truly quantum counterparts. 

\section{Funding} 

This work is supported by the the National Quantum Office, hosted in A*STAR, under the RIE 2025 AQAS projects S25Q9DA001 and S25Q9DA002, as well as the Centre for Quantum Technologies Funding Initiative (S24Q2d0009), the National Research Foundation of Singapore through the NRF Investigatorship Program (Award No. NRF-NRFI09-0010), and the Singapore Ministry of Education Tier 1 Grant RT4/23 and RG91/25. A.G. is supported by EPSRC DTP studentship EP/W524347/1 (Project 2932638) via the MADSIM CDT. T.J.E. is supported by the University of Manchester Dame Kathleen Ollerenshaw Fellowship. C.Y. is funded by Schmidt Sciences, LLC.

\bibliography{references}

\include{appendix.tex}

\end{document}

%% file: appendix.tex
\onecolumngrid
\appendix
\section*{Supplementary Materials}

\renewcommand{\thefigure}{\thesection\arabic{figure}} 
\setcounter{section}{0}
\setcounter{figure}{0}  

\setcounter{secnumdepth}{3}  
\makeatletter
\makeatother

\section{$\epsilon$-machine\label{app:e-machine}}
The $\epsilon$-machine of a stochastic process $\mathcal{P}$ is the provably memory-optimal predictive model of the process~\cite{Crutchfield2003}. Consider a bi-infinite sequence $\ldots X_{-2}X_{-1}X_0X_1X_2\ldots$, partitioned into the past $\overleftarrow{X}=\ldots X_{-2}X_{-1}$ and the future$\overrightarrow{X}=X_0X_1X_2\ldots$. The $\epsilon$-machine groups all the past $\overleftarrow{x}$ with identical futures into equivalent classes termed \textit{causal states},$\{s_i\}$, defined by the following equivalence relation:
\begin{align}
\overleftarrow{x} \sim_{\epsilon} \overleftarrow{x'} \text{ if and only if } P(\overrightarrow{X}|\overleftarrow{x}) = P(\overrightarrow{X}|\overleftarrow{x}').
\end{align}
Given a stationary stochastic process with corresponding probability distribution $P(\overleftarrow{X}, \overrightarrow{X})$, a set of causal states, $\mathcal{S} = \{s_i\},i=1,\dots, d$ for some finite $d$, can be determined, which will constitute the state space of the $\epsilon$-machine. The future probability conditioned on any causal state $s_i$ is identical to conditioning on any past $\overleftarrow{x}\in s_i$, i.e.
\begin{align}
    P(\overrightarrow{X}|s_i) = P(\overrightarrow{X}|\overleftarrow{x}_k), \forall x_k\in s_i, \quad i = 1, \dots, d
\end{align}
At each time step, the $\epsilon$-machine operates according to a set of predefined transition probabilities $T_{kj}^{(x)} = P(x, s_j|s_k)$, representing the probability that the system will output an alphabet $x\in \mathcal{X}$ and transition from causal state $s_k$ to state $s_j$. These $\epsilon$-machines are also \textit{unifilar}: given the initial causal state that the $\epsilon$-machine occupies and output $x_t$, the subsequent causal state that the $\epsilon$-machine transitions to is determined with certainty.

\section{Complexity measures for $\epsilon$-machines and quantum models\label{app:complexity-measures}}
There are two common measures of complexity associated with the $\epsilon$-machine. The first is the \emph{statistical complexity} $C_\mu$, defined as
\begin{align}
    C_\mu := H[\mathcal{S}]=-\sum_{k=1}^{|\mathcal{S}|}\pi_k \log_2 \pi_k,
\end{align}
where $H[\cdot]$ denotes the Shannon entropy, and $\pi_k$ is the stationary probability of causal state $s_k$ in the causal state distribution $\pi$ \cite{Crutchfield1989}. This quantity represents the process's entropic memory, or the minimal memory required to accurately simulate or predict the process's future, and is lower-bounded by the mutual information between the past and the future, i.e. $\mathbf{E}=I(\past{X};\future{X}) \leq C_\mu$~\cite{Shalizi2001}. The second complexity measure is the \textit{topological complexity} $d_c$, defined as:
\begin{align}
    d_c := \log_2 |\mathcal{S}|,
\end{align}
which quantifies the minimum number of bits required to store the causal states in a single-shot setting \cite{Crutchfield1989}. Correspondingly, a quantum model admits analogous complexity measures. The \emph{quantum statistical memory} $C_q$ is defined by the von Neumann entropy:
\begin{align}
    C_q := -\mathrm{Tr}[\rho\log_2\rho],
\end{align}
where $\rho=\sum_k \pi_k\ket{s_k}\bra{s_k}$ is the stationary mixed state over quantum causal states~\cite{Gu2012}. The \emph{quantum topological memory} is defined as
\begin{align}
    d_q := \log_2[\mathrm{rank}(\rho)].
\end{align}

\section{Source of quantum dimensional compression \label{app:physical-mechanism}} 
The key relation that motivates potential quantum-enhanced dimension reduction is the strict inequality $C_\mu > \mathbf{E}$, which holds for many stochastic processes~\cite{Shalizi2001}. Here, $\mathbf{E}=I(\past{X};\future{X})$ represents how much information is shared between the past and future. Naively, we would therefore expect that the minimal memory a model needs about the past of a process to predict its future would also be $\mathbf{E}$. The strict inequality, though, indicates that this is not always possible. Even the provably optimal classical model must sometimes store wasted information --- information in its memory that can never be recovered from the future. Thus, observing the entire future of a process may not allow us to retrodict precisely which state the $\epsilon$-machine was in at present:
\begin{align}
    C_\mu-\mathbf{E}=H(S\mid\future{X}).
\end{align}
This discrepancy is known in the complexity mechanics literature as \emph{crypticity}~\cite{crutchfieldTimesBarbedArrow2009}. Intuitively, one way to understand how it arises is to consider a classical $\epsilon$-machine in which two causal states have distinct conditional future distributions that nevertheless overlap. A classical machine must still allocate memory to distinguish between these states in order to know which conditional future distribution to sample from. However, seeing the entire future may not allow us to retrodict which state the machine started in, and thus not all the information allocated to storing this distinction is reflected in the future. Such a pair of causal states has more recently been referred to as a \emph{causally wasteful pair} and leads to unavoidable heat dissipation when such models are run~\cite{thompsonEnergeticAdvantagesQuantum2025}.

Quantum models can reduce this waste because they have the additional freedom to use non-orthogonal memory states. Under the mapping
$s_i\mapsto\ket{s_i}$, the model dynamics take the form
\begin{align}
    U_s\ket{s_i}\ket{0}
    =
    \sum_{x\in\mathcal{X}}\sum_k
    \sqrt{T_{ik}^{x}}\,
    \ket{s_k}\ket{x}.
\end{align}
Whenever $C_\mu>\mathbf{E}$, at least one pair of causal states cannot be perfectly distinguished by observing the future outputs. This construction then guarantees that their corresponding quantum memory states are mutually non-orthogonal, implying a strict reduction in entropic memory, such that $C_q<C_\mu$. An intuition on how these savings comes about can be gained by noting that the unitary leaves the output register and the memory entangled. Measuring the output register reproduces the required transition probabilities while collapsing the memory into the appropriate state for the next time step. As such, the quantum machine never needs to know the output symbol to make the appropriate memory transitions. 

It is important, however, to distinguish this entropic memory reduction from a reduction in the exact memory dimension. The exact quantum memory dimension and its corresponding topological memory are
\begin{align}
    D_q=\rank(\rho), \qquad d_q=\log_2D_q,
\end{align}
respectively. Non-orthogonality can lower $C_q$ without necessarily lowering $D_q$, since a set of non-orthogonal states may still span the full Hilbert space. Nevertheless, when the spectrum of $\rho$ decays rapidly, most of its
probability weight is concentrated within a much smaller subspace. In the canonical iMPS representation of the process, this spectral concentration is reflected in the corresponding bond Schmidt spectrum. It therefore becomes possible to approximate the process using an iMPS of reduced bond dimension $\tilde D$, while discarding only a small spectral weight. As shown by Yang et al.~\cite{yang2024dimension}, this observation provides a controlled route to lossy quantum dimensional compression. For a reduced bond dimension $\tilde D$, with corresponding memory cost $\tilde d=\log_2\tilde D$, the resulting distortion can be bounded in terms of $C_q/\tilde d$, as detailed in Appendix~\ref{app:quantum-dimensional-reduction}.

\section{Algorithm subroutines \label{appendix:subroutines}}
\subsection{Polar Decomposition \label{app:polar-decomposition}}
Polar Decomposition is used extensively in the construction of the unitary operators used in the algorithms. This section provides a brief explanation of the method. By Singular Value Decomposition (SVD), any $m\times n$ matrix $A$ can be decomposed as follows:
  \begin{align}
    A = U \Sigma V^\dagger,
\end{align}
where $U \in \mathbb{C}^{m\times m}$ and $V\in \mathbb{C}^{n\times n}$ are orthonormal matrices, and $\Sigma = \text{diag}(\sigma_0, \sigma_1,\dots)\in \mathbb{C}^{m\times n}$ \cite{golub2013matrix}. $U$ is referred to as the \textit{left singular matrix}, $V$ as the \textit{right singular matrix}, and the elements of $\Sigma$, i.e. $\{\sigma_0, \sigma_1\dots\}$, as the \textit{singular values}. Polar decomposition extends on SVD. Restricting $A$ to square matrices, we define two matrices
\begin{align}
    P = V\Sigma V^\dagger \\
    Q = UV^\dagger
\end{align}
where $Q$ (the \textit{orthogonal polar factor}) is unitary and $P$ (the \textit{symmetric polar factor}) is symmetric and positive semidefinite such that 
\begin{align}
    A = QP
\end{align}
The polar decomposition of any non-singular matrix is also unique \cite{golub2013matrix}. The polar decomposition is very closely related to the \textit{orthogonal Procrustes problem}, which is stated as follows: Given two data matrices $A\in \mathbb{R}^{m\times n}$ and $B\in \mathbb{R}^{m\times n}$,
\begin{align}
    \text{minimize }||A-BQ||_F , \text{ subjected to } Q^\dagger Q = I_n
\end{align}
where $||\cdot ||_F$ is the \textit{Frobenius norm} and calculated as 
\begin{align}
    ||Y||_F^2 = \sum_i\sum_j y_{ij}^2
\end{align}
If the Frobenius norm is used as the distance measure, by setting $B = I_n$, the solution to the problem gives $Q = UV^\dagger$, which is exactly the orthogonal polar factor of polar decomposition. This means that the solution of the polar decomposition can be found by finding the closest orthogonal matrix $Q$ from $A$ \cite{higham1986computing}. Therefore, $Q$ fulfils the following minimisation:
\begin{align}
    \min ||A-Q||_F
\end{align}
Applying this to our problem, if we populate any columns of $A$ with orthogonal vectors, and leave the other columns with zeros, the populated columns will be copied over into $Q$ by the nature of the minimisation of the Frobenius norm. Thus, we will get a populated unitary matrix $Q$ with the desired columns after matrix decomposition. Polar decomposition is used extensively here in the construction of the unitaries used in the quantum model. 

\subsection{Constructing the $U_s$ Operator \label{appendix:Ustep_operator}}
An approach to constructing the unitary $U_s$ involves using Kraus operators. These operators, denoted as ${K^x}, x\in \mathcal{X}$, act only on the quantum memory and satisfy the completeness relation:
\begin{align}
\sum_{x\in \mathcal{X}} (K^x)^\dagger K^x = \mathbb{I}
\end{align}
For a set of output alphabets $\mathcal{X}$, we can define a set of Kraus operators $\{K^x\}, x\in \mathcal{X}$ that describe the evolution of the quantum memory following each interaction with the ancillary qubit through the unitary matrix \cite{Yang2021}:
\begin{align}
K^x := \braket{x|U_s|0}.
\end{align}
Specifically,
\begin{align}
K^x_{ij} = \bra{i}\bra{x}U_s\ket{j}\ket{0}.
\end{align}
These Kraus operators can be obtained from the transition matrices $\{T^x\}$ through constructing a Matrix Product State (MPS) with site matrices $\sqrt{T^x_{ij}}$. The Kraus operators can be obtained through the left canonical form of this MPS \cite{yangMatrixProductStates2018}.

With the definition of the Kraus operators of the process, we can define several columns of the unitary operator $U_s$, with the remaining matrix elements being populated, for example, through a Gram-Schmidt process. These steps define the unitary construction subroutine:

\begin{algorithm}[H]
\caption{Construction of quantum models}
\begin{algorithmic}[1]
    \Procedure{QuantumModel}{$\{T^x\}$}
        \State $\{K^x\} \gets$ \Call{GetKrausOps}{$\{T^x\}$} 
        \State $U_s \gets $ \Call{CreateUStep}{$\{K^x\}$}
        \State \Return $\{K^x\}, U_s$
    \EndProcedure
    \Procedure{GetKrausOps}{$\{T^x\}$}
        \State Create MPS $A$ with site matrices $\sqrt{T^x_{ij}}$
        \State $\{K^x\}\gets $ Left canonical form of MPS $A$.
        \State \Return $\{K^x\}$
    \EndProcedure
    \Procedure{CreateUStep}{$\{K^x\}$}
    \State $n_s\gets$ dimensions of $K^x$
    \State $n_x\gets$ number of elements in $\{K^x\}$
    \State $m \gets 2^{\lceil\log_2|\mathcal{S}|\rceil}$ \Comment{round to a power of 2}
    \State $a\gets 2^{\lceil\log_2|\mathcal{X}|\rceil}$ 
    \State $U\gets $ array of size $(m\times a, m\times a)$
    \For{$x = 1$\textbf{ to } $n_x$}
        \For{$i= 1$\textbf{ to } $n_s$}
            \For{$j= 1$\textbf{ to } $n_s$}
                \State $U[x + i \cdot a, j\cdot a]\gets K^x[i,j]$ 
            \EndFor 
        \EndFor 
    \EndFor 
    \State $U_s\gets$ Polar decomposition of $U$
    \State \Return $U_s$
    \EndProcedure
    
\end{algorithmic}
\end{algorithm}

Further details on this construction method can be found in \cite{yangMatrixProductStates2018}.

\subsection{Constructing the set of initialiser matrices \label{appendix:creation_ini}}
The initialiser matrices, $\{U^i_{\mathrm{ini}}\}, i = \{0,\dots, D-1\}$ where $D$ is the number of quantum causal states, are used to initialise the memory register to the correct initial quantum causal state at the start of the simulation. Similar to the construction of the $U_s$ operator, there are two ways to construct the initialiser matrices.

\subsubsection{Using Causal States}
The first method to create the initialiser matrices involves defining the set of causal states in the computational basis, as described in \cite{Binder2018}. To create the operators necessary for the quantum circuit, we establish a set of square matrices ${A_i}, i =0, \dots, D-1$ of size $n \times n, n = 2^{\lceil \log_2 D\rceil}$ with the following structure:
\begin{align}
A_i =
\begin{bmatrix}
| & 0 & \cdots & 0\\
\ket{s_i} & 0 &\cdots & 0\\
| & 0 & \cdots & 0
\end{bmatrix},
\end{align}
where the first column is defined by the quantum causal state. Polar decomposition can be used to obtain the initialising unitary matrix $U_{\mathrm{ini}}$ of the same size as follows:
\begin{align}
A_i = U_{\mathrm{ini}}P,
\end{align}
where $P$ is some positive semidefinite Hermitian matrix. Through this method, we obtain a set of initialiser matrices $U_{\mathrm{ini}}^i, i=0,\dots, D-1$, where each causal state $\ket{s_i}$ is mapped to a unique initialiser matrix $U_{\mathrm{ini}}^i$.

\subsubsection{Using Kraus Operators}
An alternative method for generating these matrices involves the use of Kraus operators associated with the process, which can be obtained from the transition matrices as in section C.2. The unifilar property of the $\epsilon$-machine allows for the unique determination of the initial quantum causal state given a past sequence that is equal to or longer than the Markov order of the process. If a Markov order $r$ of the process and a specific past $x_{-r:0} = x_{-r}\dots x_{-2}x_{-1}$ is given, this past sequence will, by definition, be an element of a particular quantum causal state $\ket{s_i}$. This quantum causal state $\ket{s_i}$ can be produced through the repeated application of the respective Kraus operators:
\begin{eqnarray}
\ket{s_i} \propto K^{x_{-1}}K^{x_{-2}}\cdots K^{x_{-r}}\ket{\phi},
\end{eqnarray}
where $\ket{\phi}$ is in any arbitrary state. A suitable set of pasts $\{\overleftarrow{x}\}$ will allow the construction of the full set of causal states, which can be converted into initialiser matrices using the aforementioned method. The following is the subroutine \texttt{encodePast} which encodes the past into the unitary operator $U_{\mathrm{ini}}$.
\begin{algorithm}[H]
    \caption{Subroutine to encode the past into $U_{\mathrm{ini}}$}
    \begin{algorithmic}
        \Procedure{EncodePast}{$x_{-r:0}, \{K^x\}$}
        \State $m\gets \lceil \log_2 |\mathcal{S}|\rceil$ 
        \State $\mathbf{x} \gets$ leading eigenvector of one of the Kraus operators $\{K^x\}$
        \For{$i$ from $-r$ to $-1$}
            \State $\mathbf{x}\gets K^{x_i} \mathbf{x}$
        \EndFor
        \State Let $A$ be a zero matrix of size $2^m\times 2^m$
        \State $A\gets [\mathbf{x}, \mathbf{0}, \dots, \mathbf{0}]$
        \State $U_{\text{ini}}\gets $ Polar decomposition of $A$
        \State \Return $U_{\mathrm{ini}}$
        \EndProcedure
    \end{algorithmic}
\end{algorithm}

\subsection{Dimension Reduction of Quantum Models \label{app:quantum-dimensional-reduction}}

In general, the topological complexity of a process, $d_q$, upper-bounds its statistical complexity $C_q$. If the gap between $d_q$ and $C_q$ is sufficiently large, it allows for a substantial reduction in the quantum model memory dimensions without introducing significant errors. The process can be represented as an infinite Matrix Product State (iMPS)~\cite{yangMatrixProductStates2018}. We denote the bond dimension of the original iMPS by $D$ and the target bond dimension of the truncated iMPS by $\tilde{D}\leq D$. As shown in \cite{yang2024dimension}, for a given truncated dimension $\tilde{D}$, the error between the approximated quantum process $\ket{\tilde{P}}$ and the true process $\ket{P}$ is upper-bounded by $O\left(\frac{C_q}{\tilde{d}}\right)$, where $\tilde{d}=\log_2(\tilde{D})$, and the error is defined as the Quantum Fidelity Divergence Rate (QFDR) \cite{vanheckeTangentspaceMethodsTruncating2021}. A straightforward truncation can be done by keeping only the $\tilde{D}$ largest Schmidt coefficients across each bond of the iMPS and discarding the rest. The resulting reduced iMPS produces the approximate model as described in \cite{yang2024dimension}. The method adopted in this work is the improved version \cite{yang2024dimension}, where an ansatz iMPS $\tilde{A}$ of bond dimension $\tilde{D}$ is initialised and then iteratively optimized. This method yields a more stable and accurate reduced iMPS, which is used to compute the final truncated model.

\begin{algorithm}[H]
    \caption{Subroutine to truncate the quantum model}
    \begin{algorithmic}
        \Procedure{Truncate}{$\{K^x\}, \tilde{D}$}
        \State Represent $\{K^x\}$ as an iMPS $A$
        \State Initialise an ansatz iMPS $\tilde{A}$ of bond dimension $\tilde{D}$
        \State Iteratively optimize $\tilde{A}$ to maximise its overlap with target iMPS $A$ using algorithm in \cite{yang2024dimension}.
        \State Extract reduced Kraus operators $\{\widetilde{K}^x\}$ from the left canonical form of $\tilde{A}$
        \State $U_s\gets $ \Call{CreateUStep}{$\{\widetilde{K}^x\}$}
        \State \Return $\{\widetilde{K}^x\}$, $U_s$
        \EndProcedure
    \end{algorithmic}
\end{algorithm}

The main computational bottleneck in \texttt{Truncate} lies in the iterative optimization. The most computationally intensive step consists of computing the leading eigenvalues and associated left and right eigenvectors. If dense diagonalisation is used to compute the eigenvalues and eigenvectors, and $T$ iterations are required before reaching the desired error tolerance, the complexity of the algorithm is 
\begin{equation}
    \mathcal{O}\left( T(D \tilde{D})^3 \right). 
\end{equation}
If a matrix-free approach is used, the complexity of this step can be reduced to 
\begin{equation}
    \mathcal{O}\left[ TK | \mathcal{X}| \left( D^2 \tilde{D} + D \tilde{D}^2 \right) \right], 
\end{equation}
where $K$ is the number of transfer-operator applications per variational iteration.  
After convergence, the isometry defined by the iMPS must be completed to a unitary acting on a space of dimension $|\mathcal{X}|\tilde{D}$. Using a standard dense Gram--Schmidt or QR completion introduces an additional one-off worst-case cost of
\begin{equation}
    \mathcal{O}\left((|\mathcal{X}|\tilde{D})^3\right).
\end{equation}

These complexity estimates are conditional on the iteration counts $T$ and $K$, which depend on the truncation threshold, initialization, and spectral properties of the transfer operators. For fixed $T$ and $K$, the arithmetic cost is polynomial in $D$, $\tilde D$, and $|\mathcal{X}|$.

\subsection{Dimensional Reduction of Classical Models\label{app:classical-dimensional-reduction}}
To compare the performance of the dimension reduced quantum model, we construct an equivalent dimensionally reduced classical model. Given a future length $L$ that we wish to estimate, and the reduced dimension $\tilde{D}$, the optimal dimension-reduced classical model is defined as the reduced model of dimension $\tilde{D}$ which minimises the Kullback–Leibler divergence rate with the original exact classical $\epsilon$-machine \cite{yangMeasuresDistinguishabilityStochastic2020}. This reduced model is found by a brute-force search over all possible clusterings of the causal states in the original $\epsilon$-machine to $\tilde{D}$ dimensions.

\section{Gate Complexity Analysis\label{append:gate_complexity}}

For a quantum model with $D=|\mathcal{S}|$ hidden states and an alphabet size of $|\mathcal{X}|$, the memory register requires $\lceil \log_2 D\rceil$ qubits, while $\lceil \log_2 |\mathcal{X}|\rceil$ qubits are needed to store the model's output at each time step. The key component of the algorithm is the operator $\mathcal{A}$, which consists of the $U_{\mathrm{ini}}$ and $U_s$ operators. The operator $U_{\mathrm{ini}}$ prepares the memory register in the appropriate quantum causal state, and its CNOT cost is, in general, upper bounded by $\mathcal{O}(2^{\lceil \log_2 D\rceil})$ \cite{bergholmQuantumCircuitsUniformly2005, pleschQuantumstatePreparationUniversal2011}. The operator $U_s$ acts on both the memory register and the output register, involving a total of $\lceil \log_2 D\rceil+ \lceil \log_2 |\mathcal{X}|\rceil$ qubits. Its general implementation requires at most $\mathcal{O}(4^{\lceil \log_2 D\rceil+ \lceil \log_2 |\mathcal{X}|\rceil})$ CNOT gates \cite{mottonenDecompositionsGeneralQuantum2005, krolEfficientDecompositionUnitary2021, shendeSynthesisQuantumLogic2006}. Since $U_s$ needs to be applied $L$ times to simulate a sequence of $L$ time steps, the total CNOT gate complexity of implementing $\mathcal{A}$ is
\begin{align*}
    \text{CNOT}(\mathcal{A}) &\sim \mathcal{O}(L\cdot 4^{\lceil \log_2 D\rceil+ \lceil \log_2 |\mathcal{X}|\rceil} + 2^{\lceil \log_2 D\rceil}) \\
    &\sim \mathcal{O}(L\cdot 4^{\lceil \log_2 D\rceil+ \lceil \log_2 |\mathcal{X}|\rceil})\\
    &\sim \mathcal{O}(L\cdot D^2\cdot |\mathcal{X}|^2).
\end{align*}
If the quantum model is direct map from classical models using orthogonal encoding, then $|\mathcal{S}| = D_c$, and we have
\begin{align}
    \text{CNOT}(\mathcal{A}) \sim \mathcal{O}(L\cdot D_c^2\cdot |\mathcal{X}|^2).
\end{align}
If the quantum model is created using a non-orthogonal encoding, then $D = D_q \leq D_c$, and we have
\begin{align}
    \text{CNOT}(\mathcal{A}) \sim \mathcal{O}(L\cdot D_q^2\cdot |\mathcal{X}|^2).
\end{align}
Finally, if the quantum model is further compressed to $\tilde{D} < D_q$, then we have
\begin{align}
    \text{CNOT}(\mathcal{A}) \sim O(L\cdot \tilde{D}^2\cdot |\mathcal{X}|^2).
\end{align}

For the CNOT complexity of the Grover operator, it can only be defined with respect to the complexity of the binary function $f$. For a simple search protocol, the oracle $S_\chi$ and the operator $S_0$ can both be implemented using multicontrolled-$Z$ gates, which only requires a linear number of CNOT gates with respect to the number of qubits \cite{valeDecompositionMulticontrolledSpecial2023, dasilvaLineardepthQuantumCircuits2022}. As a result, their CNOT costs are $\mathcal{O}(L\cdot \lceil \log_2 |\mathcal{X}|\rceil)$ and $\mathcal{O}(L\cdot \lceil \log_2 |\mathcal{X}|\rceil+\lceil \log_2 \tilde{D}\rceil)$, respectively. Overall, the total CNOT cost of implementing the Grover operator $\mathbf{Q}$ for a search function is 
\begin{align}
    \text{CNOT}(\mathbf{Q}) \sim O(L\cdot \tilde{D}^2\cdot |\mathcal{X}|^2).
\end{align}

Assuming that the complexity of constructing $S_\chi$ and $S_0$ is simpler than constructing $\mathcal{A}$, the final CNOT complexity of the circuit is dominated by the complexity of $\mathcal{A}$. Under this assumption, using the conventional quantum amplitude estimation, to reach an absolute additive error of $\epsilon$, the number of samples required is $O(\frac{1}{\epsilon}\log \frac{1}{\delta})$, where $\delta$ is the failure rate of the algorithm \cite{Brassard2002}. Thus, to reach a certain additive error $\epsilon$ with a failure rate of $\delta$, the number of Grover operations is then
\begin{align}
    N \sim O\left(L\cdot \tilde{D}^2\cdot |\mathcal{X}|^2 \frac{1}{\epsilon}\log \frac{1}{\delta}\right)
\end{align}

\section{Maximum Likelihood Amplitude Estimation\label{app:MLAE}}
Maximum Likelihood Amplitude Estimation (MLAE) represents a family of shallow-circuit variants of standard quantum amplitude estimation. As with standard amplitude estimation, each MLAE algorithm takes in as input the algorithm operator $\mathcal{A}$ and the Grover operator $\mathbf{Q}$ and aims to output the probability estimate $\hat{p}$ of the subspace as defined by the embedded binary function $f$. Unique to the MLAE algorithm, it also takes in a \emph{schedule}, which is a set of circuits defined by $\{(m_k,N_k)\}_k$, where $m_k$ is the number of Grover operations applied in the $k$-th circuit, and $N_k$ is the number of measurement shots for the $k$-th circuit. Both $N_k$ and $m_k$ are implicit functions of the overall sample complexity $N$. The measurement results are passed into a maximum likelihood estimation subroutine to obtain the probability estimate $\widehat{p}$ of the desired sequence. Given the probability of obtaining the desired sequence $p = \sin^2\theta$, for some unknown parameter $\theta\in [0,\pi/2]$, for each of the circuit and its measurement results $\{h_k, N_k-h_k\}$, where $h_k$ is the number of times the target sequence is measured, we have a likelihood function of the following form:

\begin{align}
    L_k(h_k;\theta) \propto [\sin^{2}(\theta_{k})]^{h_k}[\cos^{2}(\theta_{k})]^{N_k-h_k}, \quad 0\leq k\leq ,M\label{eqn:individual_likelihoods}
\end{align}
where $\theta_{k} = (2m_k+1)\theta$. The unified likelihood function then is defined by
\begin{align}
    L(\mathbf{h};\theta) &= \prod_{k=0}^ML_k (h_k ;\theta).
    \label{eqn:likelihood_fn}
\end{align}
A maximum likelihood estimation is then conducted to estimate $\theta$: 
\begin{align}
    \widehat{\theta} &= \argmax_{\theta\in[0,\pi/2]}\ln L(\mathbf{h};\theta).
\end{align}
This gives the probability estimate $\widehat{p}$:
\begin{align}
    \widehat{p} = \sin^2\widehat{\theta}.
\end{align}
The two schedules used in this paper is the Linear Increment Schedule (LIS) and Exponential Increment Schedule (EIS), which is defined as follows:
\begin{itemize}
    \item LIS: $m_0 = 0, m_1 = 1, m_2=2, \dots, m_k = k$, $N_k = $ constant over all circuits.
    \item EIS: $m_0 = 0, m_1 = 1, m_2=2, m_3=4, \dots, m_k = 2^{k-1}$, $N_k= $ constant over all circuits.  
\end{itemize}
For a given sample complexity $N$ and schedule, the corresponding values of $m_k$ and $N_k$ can be determined. The exact relationship between $N, m_k$ and $N_k$ is defined by equation~\eqref{eqn:sample_complexity_equation}.

\section{Cyclic random walk\label{app:cyclic-random-walk}}
\subsection{Discretization of the cyclic random walk}
\begin{figure}[h]
    \centering
    \includegraphics[width=0.9\linewidth]{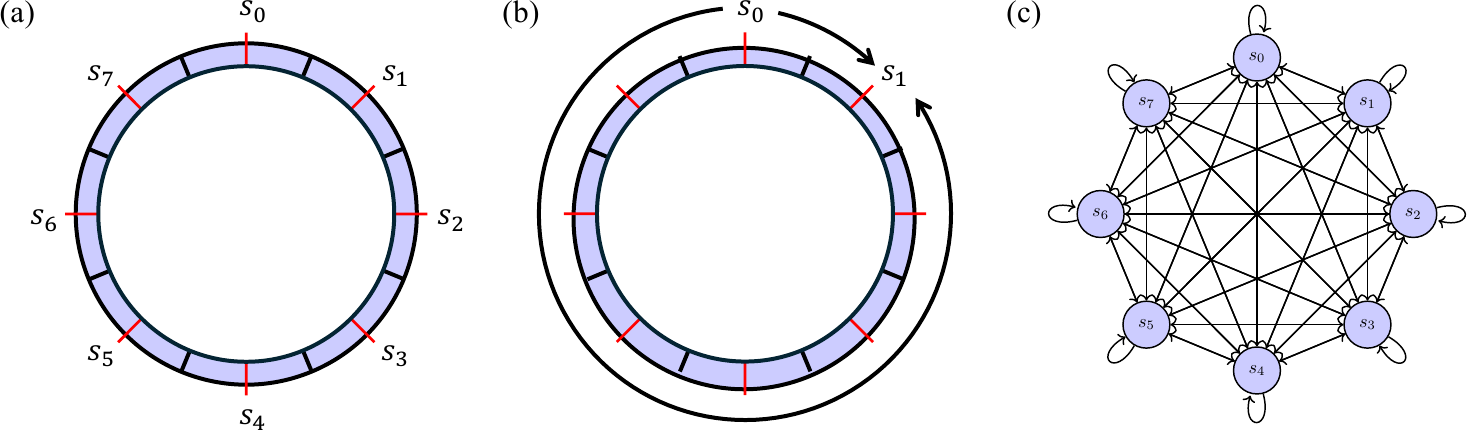}
    \caption{(a) Illustration of the discretization of the interval $[0,1)$ into eight equal parts, with the causal states $s_k$ representing the midpoint of each interval. (b) Example of the two possible paths considered for the transition from state $s_0$ to $s_1$.(c) A state machine representation of the 8-state discrete cyclic random walk. Each edge represents a transition from $s_i$ to $s_j$, with an emission of symbol $j$. The probability of transition is given in equation~\eqref{eqn:transition_prob_random_walk}.
    \label{fig:discretization_of_cyclic_walk}}
\end{figure}

For the cyclic random walk \cite{garnerUnboundedMemoryAdvantage2017}, we discretize the cycle into 8 equal intervals: $\mathcal{I}_k=\{x: \left|x-\frac{k}{8}\right| < \frac{1}{16}\}$, which give rise to 8 distinct states, $\{s_k\}_{k=0}^7$, as illustrated in Figure~\ref{fig:discretization_of_cyclic_walk}(a). The emission at each transition corresponds to the index of the resulting state. For example, a transition from $s_i$ to $s_j$ emits $j$, forming the alphabet $\mathcal{X}=\{0,1,2,\dots, 7\}$. Due to the cyclic nature of the cyclic walk, there are multiple distinct paths by which the particle may travel from one state to another, potentially looping around the circle multiple times. However, since the Gaussian distribution used for the example has a zero mean with variance $\sigma^2=0.01$, the probability of multi-cycle transitions is vanishingly small. Thus, we restrict our attention to transitions within a single cycle. Accordingly, each state transition from $s_i$ to $s_j$ is associated with only two possible paths: one clockwise and one anticlockwise around the ring. Furthermore, by symmetry, the transition probability between any two states depends only on their relative displacement along the ring. Therefore, for each state transition from $s_i$ to $s_j$, the transition probability is given by
\begin{align}
    P(x=j, s_j|s_i)=P(s_j|s_i) &= P\left(\mathcal{I}_j=\left\{\left|z-\frac{j}{8}\right|< \frac{1}{16}\right\}\left|z=\frac{i}{8}\right.\right)\\
    &= \alpha\left[\Phi\left(d_{ij} +\frac{1}{16}\right) - \Phi\left(d_{ij}-\frac{1}{16}\right)\right]+ \left[\Phi\left(d_{ij} -1+\frac{1}{16}\right) - \Phi\left(d_{ij} -1 -\frac{1}{16}\right)\right],\label{eqn:transition_prob_random_walk}
\end{align}
where 
\begin{align}
    d_{ij} = \frac{(j-i)\mod 8}{8}\in \left\{0, \frac{1}{8}, \dots, \frac{7}{8}\right\},
\end{align}
and $\alpha=1/[\Phi(15/16)-\Phi(-17/16)]$ is the re-normalisation constant to account for the discarding of multi-cycle paths, and $\Phi(x)$ is the cumulative distribution function of the Gaussian distribution. 

\subsection{Dimension reduced 4-state cyclic random walk}
\begin{figure}[h]
    \centering
    \includegraphics[width=0.3\linewidth]{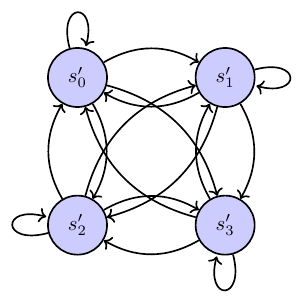}
    \caption{Illustration of the 4-state dimension reduced cyclic random walk. The transition probability and corresponding emissions are defined by equation~\eqref{eqn:reduced_transition_prob_random_walk}.}
    \label{fig:cyclic_random_walk-reduced}
\end{figure}
The optimal classical dimension reduced 4-state cyclic random walk clusters adjacent states together in the following groups: $s'_0=\{s_0, s_1\}, s'_1=\{s_2, s_3\}, s'_2=\{s_4, s_5\}, s'_3=\{s_6, s_7\}$. This gives the 4-state machine as shown in Figure~\ref{fig:cyclic_random_walk-reduced}. The state transition from $s'_i$ to $s'_j$ is then
\begin{align}
    P(x=k, s'_j|s'_i)= P(x=k, s_k|s_{2i}) + P(x=k, s_k|s_{2i+1}).\label{eqn:reduced_transition_prob_random_walk}
\end{align}
In general, the reduced quantum model exhibits lower bias compared to the reduced classical model. For certain specific events, the classical reduced model demonstrates smaller bias, but these cases are in the minority. Among the 36 events with probability between $10^{-9}$ to $10^{-8}$ used for numerical simulation, only 5 events show the quantum model producing higher bias than the classical model.

\subsection{Simulation Hyperparameters for Cyclic Random Walk\label{appendix:hyperparameters}}
For the numerical simulations on the error scaling in Figure \ref{fig:cyclic_random_walk_results}(a) and (b), the length of the future outcomes is set to $L=3$, i.e. $x_{0:3} = x_0 x_1 x_2$. The simulations are performed using the Exponential and Linear Increment Schedule (EIS and LIS, respectively), and the results are benchmarked against classical estimators. To quantify the estimation error, we compute the RMSE $\error$ (equation~\eqref{eqn:standard_error}), which is then fitted with a power-law model of the form $\error = bN^a$, where $a,b\in \mathbb{R}$ are the fitting parameters, and $N$ is the number of applications of the operator $\mathcal{A}$ (each Grover operator consist of 2 $\mathcal{A}$ operations). To calculate the sample complexity, $N$ is defined as the number of applications of the algorithm operator $\mathcal{A}$, which, for each schedule, has the following form:
\begin{align}
    N = \sum_{k=0}^M N_k(2m_k+1).\label{eqn:sample_complexity_equation}
\end{align}
This is used for comparison with classical Monte Carlo methods. The quality of the fit is evaluated via the correlation coefficient $r$. The best-fit parameters and the corresponding correlation coefficient $r^2$ values are provided in Appendix~\ref{appendix:convergence_results} (Table~\ref{tab:error_analysis}). For Figure~\ref{fig:cyclic_random_walk_results}(a) and (b), the number of shots $N_k$ for EIS and LIS is set to $200$. Each numerical experiment is repeated $1000$ times for the calculation of the RMSE. For Figure~\ref{fig:cyclic_random_walk_results}(c), among all future outcomes with $L=3$, events with probabilities between $10^{-8}$ and $10^{-9}$ (corresponding to 36 different events, details are in Appendix~\ref{app:list-of-futures}) were chosen as the target. The EIS schedule is used with $N_k=300$ shots, and $100$ independent trials were conducted to compute the RMSE. The mean RMSE of all the events are computed for the graph. Similarly, the mean classical error bound is computed from the classical error bounds of all the events. 

\subsection{List of futures used in cyclic random walk example \label{app:list-of-futures}}
For the computation of Figure~\ref{fig:cyclic_random_walk_results}(c), starting from the past $x_{-3:0}=000$, and estimating three steps into the future, 36 different future events have an occurrence probability between $10^{-9}$ to $10^{-8}$. These events are listed as follows: 
\begin{align}
    x_{0:3}=\{037, 0 4 1, 0 4 7, 0 5 1, 1 4 0,  1 5 0,1 5 2, 1 6 2,3 2 6, 3 3 7,3 4 0, 3 7 0,\nonumber\\
    3 7 6,3 7 7,4 1 0, 4 1 1,4 1 2,4 3 0,4 3 6,4 4 1,4 4 7,4 5 0,4 5 2,4 7 0,\nonumber\\ 4 7 6,4 7 7,
    5 1 0,5 1 1,5 1 2,5 4 0,5 5 1,5 6 2,7 2 6,7 3 0,7 3 6,7 4 0\} 
\end{align}

For the computation of Figure~\ref{fig:16state}, we utilize the same past of $x_{-3:0}=000$, and estimate the probabilities two steps into the future, which corresponds to 12 different future have an occurrence probability between $10^{-9}$ to $10^{-8}$. These events are listed as follows:
\begin{align}
    x_{0:2} = \{ &[5,13], [6, 13], [6, 15], [7, 1], \nonumber\\
    &[7, 13], [8, 3], [8, 13], [9, 3], \nonumber\\
    &[9, 15], [10, 1], [10, 3], [11, 3] \}
\end{align}

\subsection{Improving Simulation Accuracy through Dimensional Reduction on Noisy Processors  \label{app:noisy_proc}} 

On an ideal quantum hardware, dimensional reduction trades a small, controlled truncation bias for a significant reduction in the required qubit count and circuit complexity. On noisy processors, however, this resource reduction provides an additional practical advantage: decreasing the size of the unitary required to simulate the process may also substantially reduce the gate count and gate error accumulation. The corresponding reduction in hardware-induced error can outweigh the truncation bias, allowing the truncated model to achieve a lower estimation error than the untruncated model.

In this section, we illustrate an example where a quantum computer with a noise profile comparable to that of contemporary NISQ devices can perform better when using our dimensional-reduction method. We utilize the same eight-state cyclic random walk as the reference model, and the corresponding four-state memory-reduced model. As demonstrated in the paper, the memory reduction introduces only a small amount of bias. We consider a quantum device with a CNOT depolarizing error rate of $7.9\times 10^{-4}$ and a single-qubit gate depolarizing error rate of $2.5\times 10^{-5}$, chosen to be comparable to the performance of the Quantinuum Helios processor~\cite{HeliosQuantinuumsQuantum}.  We take $P(Y)$ to be a zero-mean Gaussian distribution with variance $\sigma^2=0.01$.

Using these parameters, we benchmark the ability of the two models to estimate the probability of the one-step future sequence $x_{0:1}=1$, conditioned on the past $\overleftarrow{x}=000$. The true probability is $ \approx 0.235589$. For each model, we draw $N_{\mathrm{shots}}=10,000$ samples, and repeat the experiment across $10$ independent trials before computing the mean. The untruncated model yields a mean estimate of $\hat{p}=0.1372(28)$, whereas the truncated model yields $\hat{p}=0.1929(26)$. The corresponding absolute errors are $0.098$ and $0.043$, respectively, amounting to a reduction in estimation error of more than $50\%$.

\begin{figure}[htbp]
    \centering
    \includegraphics[width=0.5\linewidth]{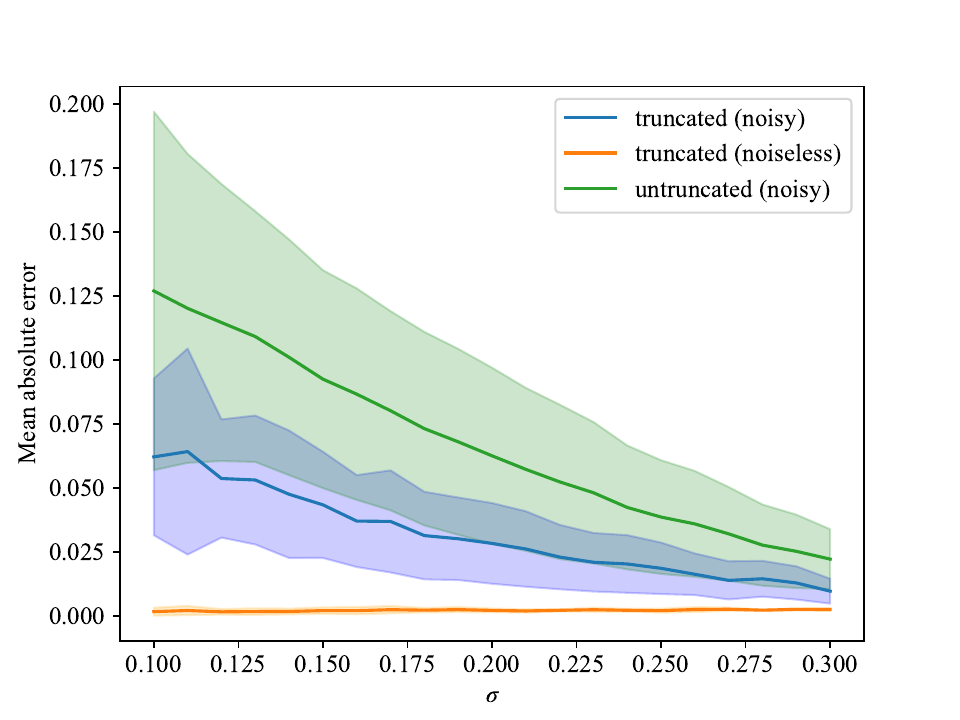}
    \caption{Mean absolute errors of the noisy truncated and untruncated models for different values of $\sigma$, with the noiseless truncated model shown as a baseline. The truncated model has a lower mean absolute error than the untruncated model throughout the tested parameter range.}
    \label{fig:mean_absolute_error_noisy}
\end{figure}

We extend this analysis to all eight possible one-step outputs. Averaged over these outputs, the untruncated model yields a mean absolute error of $0.126$, whereas the truncated model yields a mean absolute error of $0.0679$. Thus, the approximately $50\%$ reduction in estimation error persists across the full output alphabet. Finally, we perform a parameter sweep over different instances of the cyclic random walk by varying $\sigma$ from $0.1$ to $0.3$. As shown in Figure~\ref{fig:mean_absolute_error_noisy}, the reduction in estimation error persists throughout the tested parameter range.

\subsection{Finer Discretization and the Continuum Limit}
\label{app:finer-discrete}

Our main analysis has focused on a target process obtained by discretizing the cyclic random walk to $k=3$ bits of precision, corresponding to $N=2^k=8$ discrete states. This choice was made for proof-of-principle purposes. As the discretization is made finer by increasing $k$, the resulting target process approximates the underlying continuous-valued cyclic random walk ever more closely. It is therefore natural to ask whether the benefits of quantum memory reduction persist as the discretization distortion is reduced.

To formalize this, let $P^{(k)}$ denote the target stochastic process obtained from the $k$-bit discretization of the cyclic random walk, with $N=2^k$ states. An exact classical model of $P^{(k)}$ therefore requires $k$ bits of memory. Meanwhile, let $\tilde P^{(k)}$ denote an approximation generated by a dimensionally reduced quantum model with memory dimension
\begin{align}
    \tilde D = 2^n,
\end{align}
such that only $n<k$ memory qubits are required. The iMPS framework underlying our dimensional-reduction method provides a useful theoretical bound on the distortion introduced by this reduction. Let $\ket{P^{(k)}}$ and $\ket{\tilde P^{(k)}}$ denote the corresponding quantum samples. Their distortion can be quantified by the quantum fidelity divergence rate (QFDR),
\begin{align}
R_F\left(\ket{P^{(k)}},\ket{\tilde P^{(k)}}\right) := -\lim_{L\rightarrow\infty} \frac{1}{2L} \log_2 F\left( \ket{P^{(k)}_L}, \ket{\tilde P^{(k)}_L} \right),
\end{align}
where
\begin{align}
F(\ket{\psi},\ket{\phi})=|\braket{\psi|\phi}|.
\end{align}
Here, $\ket{P^{(k)}_L}$ and $\ket{\tilde P^{(k)}_L}$ denote the respective length-$L$ quantum samples. A low QFDR also implies a low statistical fidelity divergence rate between the corresponding stochastic processes. Let
\begin{align}
C_q^{(k)}=-\mathrm{Tr}\left[\rho^{(k)}\log_2\rho^{(k)}\right]
\end{align}
denote the quantum statistical complexity of $P^{(k)}$, where $\rho^{(k)}$ is the stationary state of the quantum memory. Ref.~\cite{yang2024dimension} shows that, when the available quantum memory is restricted to dimension $\tilde D=2^n$, there exists a reduced quantum model satisfying
\begin{align}
R_F\left(\ket{P^{(k)}},\ket{\tilde P^{(k)}}\right)\leq \mathcal{O}\left(\frac{C_q^{(k)}}{n}\right)=
\mathcal{O}\left(\frac{C_q^{(k)}}{\log_2\tilde D}\right).
\label{eq:finer-discretization-bound}
\end{align}
Thus, when the quantum statistical complexity is much smaller than the number of memory qubits available, significant dimensional reduction can be achieved while introducing only small distortion.

This observation is particularly relevant to the cyclic random walk considered here. For cyclic random walks subject to Gaussian noise of fixed nonzero width, the quantum statistical complexity remains bounded as the discretization becomes progressively finer~\cite{garnerUnboundedMemoryAdvantage2017}. That is, there exists some finite constant $C_\ast$ such that
\begin{align}
C_q^{(k)} \leq C_\ast
\end{align}
for arbitrarily large $k$, even though an exact classical representation requires $2^k$ memory states, or $k$ bits. We emphasize that this boundedness concerns the \emph{entropic} quantum memory $C_q^{(k)}$; the Hilbert-space dimension required for an exact quantum model need not itself remain bounded. It is precisely this separation between information content and Hilbert-space dimension that makes quantum dimensional reduction useful.

We can now consider increasing the memory available to the reduced quantum model much more slowly than the precision of the target process. For example, choose
\begin{align}
n(k)=\left\lceil\sqrt{k}\right\rceil,
\qquad
\tilde D(k)=2^{n(k)}.
\end{align}
The reduced quantum model then uses an asymptotically vanishing fraction of the memory qubits required to represent the $k$-bit target process exactly,
\begin{align}
\frac{n(k)}{k} \longrightarrow 0
\qquad
\text{as}
\qquad
k\longrightarrow\infty.
\end{align}
At the same time, Eq.~\eqref{eq:finer-discretization-bound} gives
\begin{align}
R_F\left(\ket{P^{(k)}},\ket{\tilde P^{(k)}}\right)\leq\mathcal{O}\left(\frac{C_\ast}{\sqrt{k}}\right)\longrightarrow 0.
\end{align}
Thus, the additional distortion introduced by quantum memory reduction can vanish even while the number of memory qubits used by the reduced model grows parametrically more slowly than the precision $k$ of the target process.

Meanwhile, increasing $k$ also makes the target process $P^{(k)}$ an increasingly fine discretization of the underlying continuous random walk~\cite{garnerUnboundedMemoryAdvantage2017}. We can therefore simultaneously reduce both sources of approximation: the discretization distortion between the underlying continuous process and $P^{(k)}$, and the memory-reduction distortion between $P^{(k)}$ and $\tilde P^{(k)}$. Importantly, this can be achieved while using far fewer than $k$ qubits of quantum memory. Thus, finer discretization does not force the distortion introduced by quantum memory reduction to increase. Together with our numerical results for both the $k=3$ and $k=4$ target processes, where the reduced quantum models outperform classical baseline models subject to the same memory constraints, this provides strong evidence that the quantum advantage observed here is not specific to the particular discretization precision used in our original proof-of-principle example.

\section{Additional examples \label{app:examples}}

\subsection{Autoregressive Model \label{app:ar2-model}}
\begin{figure}[h]
    \centering
    \includegraphics[width=0.4\linewidth]{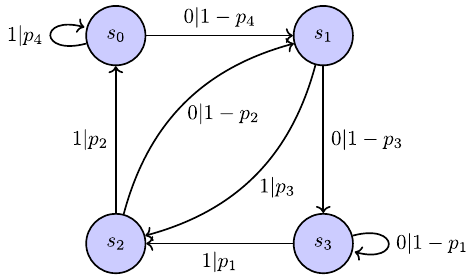}
    \caption{$\epsilon$-machine of the discretised AR(2) process. The value of $p_i, i \in \{1,2,3,4\}$ are parameterised by $\kappa_1$ and $\kappa_2$. The exact relationships are shown in Section~\ref{appendix:ar2_discrete}.}
    \label{fig:e-machine ar2}
\end{figure}

The autoregressive (AR) model is a simple model frequently used to represent time series processes, including security prices in finance \cite{shumwayTimeSeriesAnalysis2000}. Consider the autoregressive model with order 2, AR(2), where the output $x_t$ is time $t$ obeys
\begin{align}
    X_t = \kappa_1 X_{t-1} + \kappa_2 X_{t-2} + \xi_t, \label{eqn:ar2}
\end{align} 
$\kappa_1, \kappa_2 \in \mathbb{R}$ are the set of weights on the previous time steps, and $\xi_t$ is the white noise at time $t$. This gives a process with Markov order $2$. Since the original AR(2) model takes values in $\mathbb{R}$, we first discretise it such that $X_t\in \{0,1\}$. This generates the set of transition matrices, which are then used to construct the quantum model. The $\epsilon$-machine for the discretised process is shown in Figure~\ref{fig:e-machine ar2}, and the discretisation procedure is provided in Section~\ref{appendix:ar2_discrete}. 

\begin{figure*}[t]
    \centering
    \includegraphics[width=0.9\linewidth]{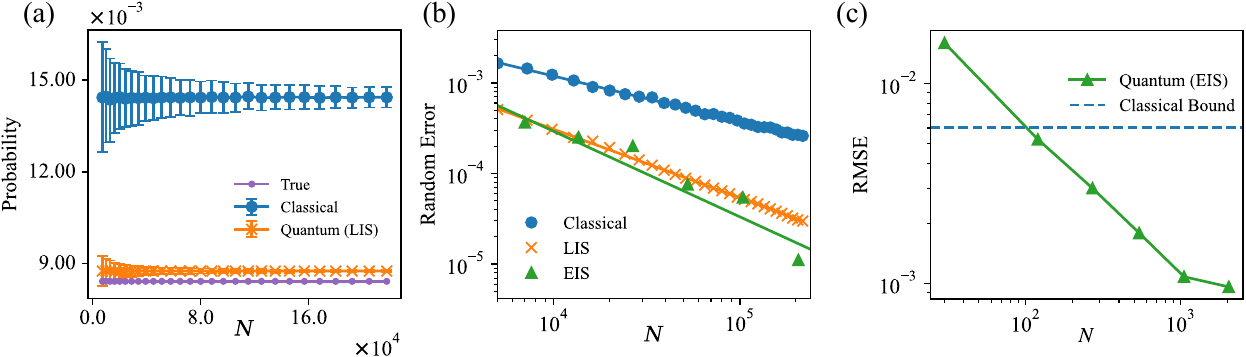}
    \caption{(a) Trend of the convergence of the mean estimate of the algorithm over 1000 independent trials. The convergence of the classical algorithm is plotted for comparison, with the error bar corresponding to the 95th and 5th percentile. (b) Error trend of the different schedules as $N$ increases. The numerical values of the fitting parameters and the corresponding correlation coefficients $r^2$ are given in Table~\ref{tab:error_analysis} in the Appendix. (c) Average error for estimating the probability of three events $x_{0:4}=\{1001, 1100, 1101\}$, conditioned on $x_{-3:-1}=000$ for the dimension reduced 3-state quantum model for different number of samples, compared with the classical bound for a classical dimension reduced 3-state machine.}
    \label{fig:ar2_results}
\end{figure*}

We utilised our algorithm to estimate the probability of a specific event in this process. The parameters $\kappa_1$ and $\kappa_2$ are set to $0.5$ and $0.4$, respectively. Starting from state $s_0$, the event chosen is $x_{0:4}=1001$. with probability $P(x_{0:4}=1001|s_0) \approx 0.00842$. The results are displayed in Figure \ref{fig:ar2_results}, where a quantum speedup over the classical algorithm is observed.

Furthermore, we apply the method in \cite{yang2024dimension} to reduce the dimension of the discretised AR(2) model from four states to three states. This creates an approximate model that will no longer be able to capture the full statistics of the process. We compare both the effect of truncation on the accuracy of probability estimation for both the quantum and classical models. The truncated quantum model was used to estimate the probability of events with probability less than $10^{-2}$, starting from a past $x_{-3:0}=000$. This corresponds to 3 different events: $x_{0:4}=\{1001, 1100, 1101\}$. For the classical comparison, a search through all possible classical three-state machine is conducted as per \cite{Yang2021}, and a classical bound is computed. The results (Figure~\ref{fig:ar2_results}c) show the average estimation error of these 3 events compared to the average classical bound. The error of the quantum estimate falls below the average classical limit even with just $N\sim 10^2$ samples, highlighting the benefit of the quantum dimension reduction. 

\subsection{Perturbed Coin\label{appendix:perturbed_coin}} 
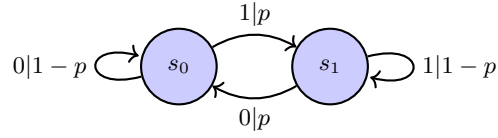
\begin{figure}[h]
\begin{tikzpicture}[->,node distance =2cm, thick]
\tikzstyle{every state}=[fill={blue!20}, minimum size=1cm]]
\node[state] (1) {$s_0$};
\node[state, right of=1] (2) {$s_1$};
\path   (1) edge[above, bend left] node{$1|p$} (2)
        (1) edge[loop left] node{$0|1-p$} (1)
        (2) edge[loop right] node{$1|1-p$} (2)
        (2) edge[below, bend left] node{$0|p$} (1);
\end{tikzpicture}

\caption{\label{fig:perturbed_coin} $\epsilon$-machine of the perturbed coin process. The process has two causal states, and it is characterised by the transition probability $p$ between each causal state.}
\end{figure}

The classical $\epsilon$-machine for the perturbed coin process is depicted in Figure \ref{fig:perturbed_coin}.

\begin{figure}[h]
    \centering
    \includegraphics[width=0.6\linewidth]{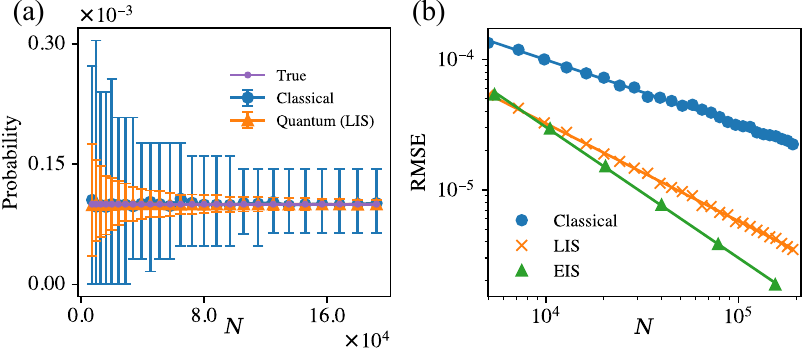}
    \caption{(a) Trend of convergence for the mean estimate of the algorithm over 1000 samples for the perturbed coin process, with the transition probability $p=0.9$ and the initial state set to $s_0$. For comparison, the convergence of the classical algorithm is plotted. The error bars represent the 95th and 5th percentiles of probability estimates in the sample of 1000. (b) Error trend of different schedules as the number of shots increases.}
    \label{fig:perturbed_coin_results}
\end{figure}

The algorithm is utilized to estimate the future sequence $x_{0:4} = 0000$ from the initial state $s_0$. With the transition probability set at $p=0.9$, this results in a theoretical probability of $P(0000|s_0) = 0.0001$. The convergence of the error is shown in Figure~\ref{fig:perturbed_coin_results}.

\subsection{Nemo Process\label{appendix:nemo_process}}

\begin{figure}[h]
\begin{tikzpicture}[->, node distance =3cm, thick,auto]
\tikzstyle{every state}=[fill={blue!20}, minimum size=1cm]
\node[state] (1) {$s_1$};
\node[state, below left of=1] (2) {$s_0$};
\node[state, below right of=1] (3) {$s_2$};

\path   (1) edge[above right] node{$1|1$} (3)
        (2) edge[left, loop left] node{$0|p$} (2)
        (2) edge[above left] node{$1|1-p$} (1)
        (3) edge[below, bend left] node{$0|0.5$} (2)
        (3) edge[above, bend right] node{$1|0.5$} (2);
\end{tikzpicture}
\caption{\label{fig:nemo_process} $\epsilon$-machine of the Nemo process. The single variable in this process is $p$, which determines the probability of the system staying in the state $s_0$.}
\end{figure}
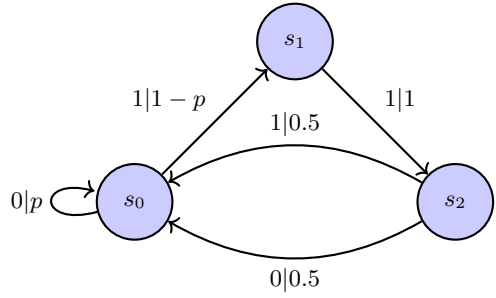

The Nemo process comprises three causal states, necessitating two qubits to store the memory in the quantum model. The first causal state, $s_0$, has a probability $p$ of emitting a 0 and remaining in the same state, and a probability of $1-p$ of emitting a 1 and transitioning to the second causal state, $s_1$. The state $s_1$ transitions to $s_2$ with a probability of 1, emitting a 1 in the process. Lastly, from $s_2$, there is an equal probability of emitting a 1 or 0 before the system transitions back to $s_0$. The $\epsilon$-machine is depicted in Figure \ref{fig:nemo_process}.

\begin{figure}[htbp]
    \centering
    \includegraphics[width=0.6\linewidth]{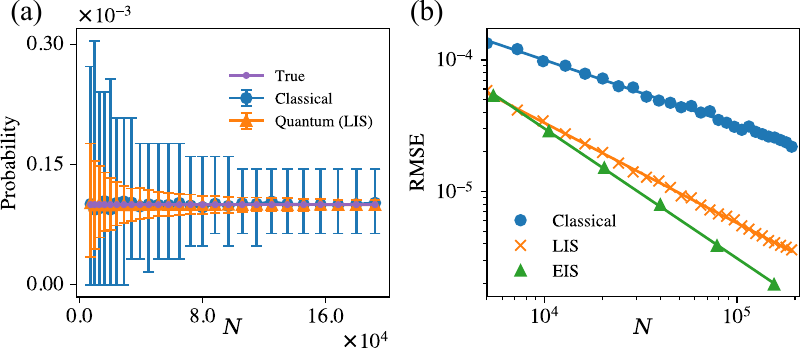}
    \caption{(a) Convergence trend of the mean estimate of the algorithm over 1000 samples. For comparison, the convergence of the classical algorithm is plotted. The error bar represents the 95th and 5th percentiles of the probability estimates across the 1000 samples. (b) Error trend of different schedules as the number of shots increases.}
    \label{fig:nemo_results}
\end{figure}

The algorithm was applied to estimate the probability of the future sequence $x_{0:4}=0000$ in the Nemo process, with the quantum model initialized in the quantum causal state $s_0$. With the parameter $p=0.1$, the true probability of this future sequence is $P(x_{0:4} = 0000|s_0) = 0.0001$. The results are presented in Figure \ref{fig:nemo_results}.
\FloatBarrier

\subsection{Dual Poisson process\label{app:dpp}} 
\begin{figure}[ht]
\begin{tikzpicture}[->, node distance =2.2cm, thick]
\tikzstyle{every state}=[fill={blue!20}, minimum size=0.9cm]]
\node[state] (1) {$s_0$};
\node[state, right of=1] (2) {$s_1$};
\node[state, right of=2] (3) {$s_k$};
\node[state, right of=3] (4) {$s_{k+1}$};

\draw   (1) edge[above] node{$0|\frac{\Phi(1)}{\Phi(0)}$} (2)
        (1) edge[loop above] node{$1|1-\frac{\Phi(1)}{\Phi(0)}$} (1)
        (2) edge[above, bend left] node{} (1)
        (2) edge node[fill=white]{$\cdots$}(3)
        (3) edge[below right, bend left] node{$1|1-\frac{\Phi(k+1)}{\Phi(k)}$} (1)
        (3) edge[above] node{$0|\frac{\Phi(k+1)}{\Phi(k)}$} (4);
\end{tikzpicture}
\caption{\label{fig:dual_poisson-e-machine} $\epsilon$-machine of a dual Poisson process. The system begins in state $s_0$ and transitions to a new state whenever a 0 is emitted, but it resets to $s_0$ when a 1 is emitted.}
\end{figure} 

The dual Poisson process \cite{Elliott2020} is a highly non-Markovian process that is parameterised by $q, r_1, r_2, 0\leq q, r_1, r_2\leq 1$, which defines the so-called \textit{survival probability} $\Phi(k)$ as follows:
\begin{eqnarray}
\Phi(k) = qr_1^k + (1-q)r_2^k
\label{eqn:survival_prob}
\end{eqnarray}
The process's $\varepsilon$-machine is depicted in Figure \ref{fig:dual_poisson-e-machine}. It was shown in \cite{Elliott2020} that the $\varepsilon$-machine to generate a contiguous string of $k$ $0$s requires an infinite number of causal states as $k$ goes to infinity. This contrasts sharply with the corresponding quantum model, which only requires one qubit to simulate this process \cite{Elliott2020}.

\begin{figure}[htbp]
    \centering
    \includegraphics[width=0.6\linewidth]{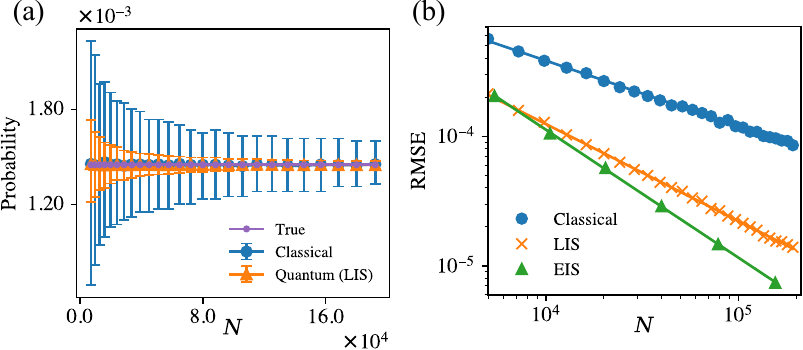}
    \caption{(a) Convergence trend of the mean estimate of the algorithm over 1000 samples. For comparison, the convergence of the classical algorithm is plotted. The error bar represents the 95th and 5th percentiles of the probability estimates across the 1000 samples. (b) Error trend of different schedules as the number of shots increases.}
    \label{fig:dual_poisson_results}
\end{figure}

We employ our quantum probability estimation algorithm to estimate the probability of the sequence $x_{0:4} = 0000$, from the initial state $s_0$. With parameters $q=0.1, r_1=0.1$ and $r_2=0.2$, the theoretical probability is $P(x_{0:4}=0000|s_0) = 0.00145$. The findings are shown in Figure~\ref{fig:dual_poisson_results}. The quantum algorithm exhibits a speedup over the classical method, with the quantum algorithm’s standard error decreasing faster with increasing Grover iterations than with the corresponding number of shots in the classical algorithm. The observed speedup factors align with the predicted error scaling.

\subsection{Table of numerical results  \label{appendix:convergence_results}}

\begin{table*}[h]
   \centering
   \begin{tabular}{c|c|c|c|c|c|c}
       \hline
       \hline
      \textbf{Schedule} & \multicolumn{2}{c|}{\textbf{Classical}} & \multicolumn{2}{c|}{\textbf{LIS}} & \multicolumn{2}{c}{\textbf{EIS}} \\
      \hline
      \textbf{Process} & $a$ & $r^2$ & $a$ & $r^2$ & $a$ & $r^2$ \\
      \hline
      \textbf{Perturbed Coin} &$-0.4946\pm 0.0055$& $0.9969$&$-0.7498\pm 0.0042$&$0.9992$& $-1.004\pm 0.011$&$0.9995$ \\
      \textbf{Nemo Process} &$-0.4962\pm 0.0067$&$0.9955$&$-0.7591\pm 0.0044$&$0.9991$&$-0.9847\pm 0.0097$&$0.9996$ \\
      \textbf{Dual Poisson} &$-0.5066\pm 0.0043$&$0.9982$&$-0.7489\pm 0.0032$&$0.9996$&$-0.9871\pm 0.0057$&$0.9999$ \\
      \textbf{AR(2) (exact)} &$-0.4906\pm 0.0053$& $0.9970$ & $-0.7476\pm 0.0040$ & $0.9993$ & $-0.875 \pm 0.064$ & $0.9792$\\
      \textbf{AR(2) (truncated)} &$-0.4978\pm 0.0034$& $0.9987$& $-0.7553\pm 0.0032$& $0.9995$&$-0.944\pm 0.077$& $0.9560$ \\
      \textbf{Cyclic Random Walk (exact)} & $-0.5037\pm 0.0045$ & $0.9979$ & $-0.7529\pm 0.0034$ & $0.9995$ & $-0.995\pm 0.014$ & $0.9994$\\
      \textbf{Cyclic Random Walk (truncated)} &$-0.5024 \pm 0.0038$& $0.9985$ & $-0.7682\pm 0.0086$& $0.9967$& $-1.085\pm 0.076$& $0.9762$ \\
      \hline
      \textbf{Average} & $0.4988$ && $0.7545$ && $0.982$ &\\
         \hline
         \hline
   \end{tabular}
    \caption{Table of the index of the best-fit error scaling for all the processes, with the corresponding $r^2$ correlation value to indicate fit. The exact reference for AR(2) and cyclic random walk are the discretized version of the continuous process (the 4-state machine for AR(2) and 8-state machine for cyclic random walk). Here, the errors denote the estimated standard deviation of the parameter estimate, computed using the function \texttt{curve\_fit} from the \texttt{scipy.optimize} package in Python.}
    \label{tab:error_analysis}
\end{table*}

\FloatBarrier

\section{Discretization of AR(2) process \label{appendix:ar2_discrete}}
The $AR(2)$ process over real values is defined as:
\begin{align}
    X_t = \phi_1 X_{t-1} + \phi_2 X_{t-2} + \epsilon_t,
\end{align}
where $\phi_1,\phi_2 \in \mathbb{R}$ and
\begin{align}
    \epsilon_t \sim N(0, \sigma_\epsilon^2)
\end{align}
for some $\sigma_\epsilon^2 > 0$. The output space is discretized into two different output symbols:
\begin{align}
    X^{(D)}_t =
    \begin{cases}
    -1 &\text{ if } X_t < 0\\
    1 &\text{ if } X_t \geq 0,
    \end{cases} \label{eqn: discretization}
\end{align}
Since the process has a Markov order of 2, the full set of states is given as $\{(-1)(-1), (-1)(1), (1)(-1), (1)(1)\}$.
The output symbols from each of these states can be computed exactly:
\begin{align}
    (-1)(-1) : X_t &= \phi_1 \cdot -1 + \phi_2 \cdot -1 + \epsilon_t \nonumber\\
                    &= \epsilon_t - \phi_1 - \phi_2 \label{eqn:undiscretized_AR(2)_1}\\
    (-1)(1) : X_t &= \phi_1 \cdot 1 + \phi_2 \cdot -1 + \epsilon_t \nonumber\\
                    &= \epsilon_t +\phi_1 - \phi_2 \label{eqn:undiscretized_AR(2)_2} \\
    (1)(-1): X_t &= \phi_1 \cdot -1 + \phi_2 \cdot 1 + \epsilon_t \nonumber\\
                    &= \epsilon_t - \phi_1 + \phi_2 \label{eqn:undiscretized_AR(2)_3}\\
    (1)(1): X_t &= \epsilon_t + \phi_1 + \phi_2 ,\label{eqn:undiscretized_AR(2)_4}
\end{align} 
We define:
\begin{align}
    p_1 &= P(\epsilon_t > \phi_1 + \phi_2)\\
    p_2 &= P(\epsilon_t > -\phi_1 + \phi_2)\\
    p_3 &= P(\epsilon_t > \phi_1 -\phi_2)\\
    p_4 &= P(\epsilon_t > -\phi_1-\phi_2),
\end{align}
With this, the definition of the discretization process defined in equation \eqref{eqn: discretization} and the dynamics of the undiscretized process defined by \eqref{eqn:undiscretized_AR(2)_1} -- \eqref{eqn:undiscretized_AR(2)_4} gives the following hidden Markov model:

\begin{figure}[h]
    \centering
    \includegraphics[width=0.5\linewidth]{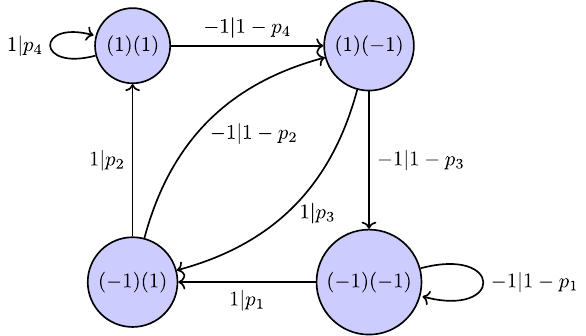}
    \label{fig:ar2_intermediate_state_machine}
\end{figure}
\FloatBarrier
We can then rewrite each state with the causal state notation $s_i$, for $i = 0,1,2,3$, and map the output values from $\{-1,1\}$ to $\{0,1\}$ for ease of coding:

\begin{figure}[h]
    \centering
    \includegraphics[width=0.45\linewidth]{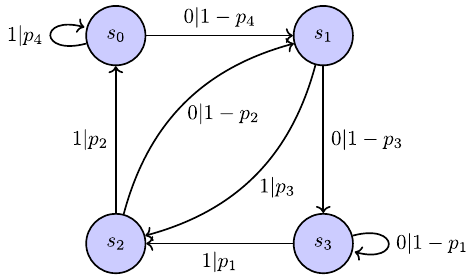}
    \label{fig:e-machine-ar2-app}
\end{figure}

Thus, we have the following transition matrices
\begin{align}
    T^0 =
    \begin{bmatrix}
    0 & 1-p_4 & 0 & 0\\
    0 & 0 & 0 & 1-p_3\\
    0 & 1-p_2 & 0 & 0\\
    0 & 0 & 0 & 1-p_1
    \end{bmatrix}, \quad
    T^1 = 
    \begin{bmatrix}
    p_4 & 0 & 0 & 0\\
    0 & 0 & p_3 & 0\\
    p_2 & 0 & 0 & 0\\
    0 & 0 & p_1 & 0
    \end{bmatrix}
\end{align}
where for $i,j \in \{0,1,2,3\}, k \in \{0,1\}$, $T^k_{ij}$ corresponds to $P(x_t=k, s_t = s_j | s_{t-1}=s_i)$.